\documentclass{elsarticle}

\usepackage[a4paper]{geometry}
\usepackage{comment}
\usepackage[textwidth=8em,textsize=small]{todonotes}
\usepackage{amsmath}
\usepackage{enumitem} %for nosep
\usepackage{natbib}
\usepackage{graphicx} %for resizebox
\usepackage{xcolor}
\usepackage{amsfonts}
\usepackage{pdfpages}
\usepackage{bm} %bold symbols

\usepackage{multicol}
\usepackage{multirow, rotating}
\usepackage{tabularx}
\usepackage[bookmarks=false, pdfauthor=author]{hyperref} %for arXiv submission

\usepackage{subfig}
\graphicspath{{figures/}}

\usepackage{algorithm,algpseudocode,amsmath}

\makeatletter
\newcounter{phase}[algorithm]
\newlength{\phaserulewidth}
\newcommand{\setphaserulewidth}{\setlength{\phaserulewidth}}

\makeatother

\setphaserulewidth{.7pt}

\biboptions{authoryear}

\begin{document}

\begin{frontmatter}

\title{A Two-Stage Cox Process Model with Spatial and Nonspatial Covariates}

\author[mymainaddress]{Claire Kelling\corref{mycorrespondingauthor}}%\fnref{myfootnote}
\cortext[mycorrespondingauthor]{Corresponding author}
\ead{cekelling@gmail.com}
\ead[address]{cekelling@gmail.com}
\fntext[myfootnote]{Department of Statistics,
326 Thomas Building
University Park, PA 16802}

\address[mymainaddress]{Department of Statistics,
326 Thomas Building
University Park, PA 16802, United States}

\author[mymainaddress]{Murali Haran}%\fnref{myfootnote}

\begin{abstract}
Rich new marked point process data allow researchers to consider disparate problems such as the factors affecting the location and type of police use of force incidents, and the characteristics that impact the location and size of forest fires. We develop a two-stage log Gaussian Cox process that models these data in terms of both spatial (community-level) and nonspatial (individual or event-level) characteristics; both types of covariates are present in the examples we consider and are not easy to incorporate via existing methods. Via simulated and real data examples we find that our model is easy to interpret and flexible, accommodating multiple types of marks and multiple types of spatial covariates. In the first example we consider, our approach allows us to study the impact of community-level socioeconomic features such as unemployment as well as event-level features such as officer tenure on force used by police, illustrated through simulated examples. In our second example we consider factors that impact the locations and severity of forest fires from the Castilla-La Mancha region of Spain between 2004-2007.

\vspace{6pt}

\noindent Declaration of interest: none
\end{abstract}


\end{frontmatter}

\section{Introduction}

Rich marked point process data are increasingly available in a number of domains. Such data enable researchers to answer interesting new questions. However, these data often pose new statistical and computational challenges. We consider two motivating examples: police use of force where the mark is the level of force used by police, and forest fires, where the mark is the amount of burned area caused by each fire.  In addition to understanding the nature and driving factors of individual use of force incidents or forest fires, it is also critical for scientists, policymakers, and the public to develop a data-driven, systematic understanding of the important factors at play in driving outcomes generally.

\begin{figure}[ht]
    \centering
    \includegraphics[width = 0.9\textwidth]{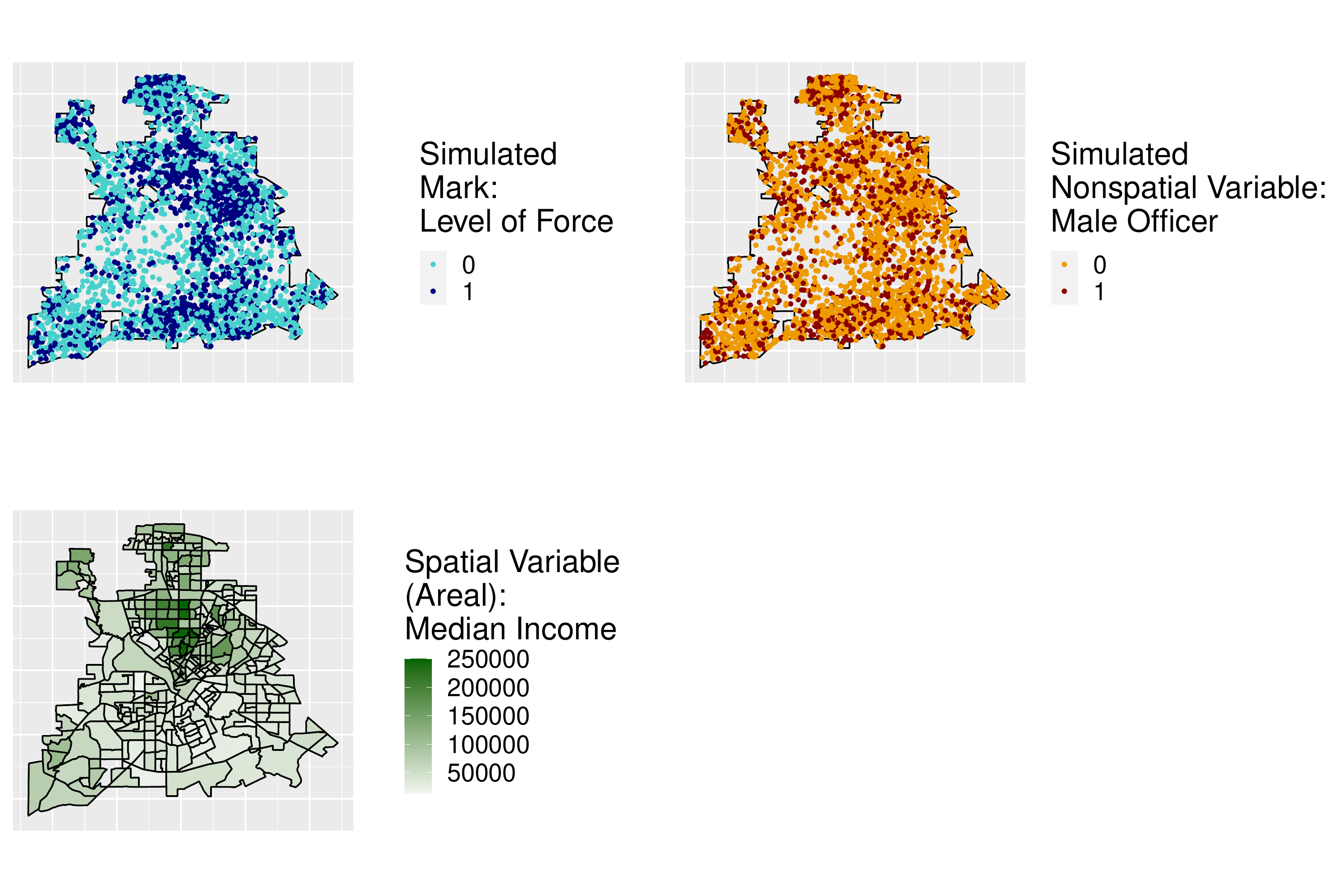}
    \caption{Simulated Police Use of Force Data}
    \label{fig:uofdata}
\end{figure}

The first motivating dataset we describe is police use of force events in the city of Dallas, Texas. Fully understanding police use of force requires a model for a point process that has multiple types of predictors, including both spatial and nonspatial variables. Examples of spatial variables that may impact police use of force are neighborhood-level characteristics such as the unemployment rate or neighborhood diversity. Nonspatial variables include both individual-level characteristics, such as officer or citizen race and officer training, and event-level characteristics, such as presence of citizen resistance or if the officer or citizen was injured. Our modeling framework determines the potential impact of these variables on the locations of events and on the level of force used by police (the mark). Examples of simulated marks and nonspatial variables following the structure of use of force data from Dallas, Texas in plotted in Figure \ref{fig:uofdata}, with spatial variables from the American Communities Survey plotted at the bottom.

\begin{figure}[ht]
    \centering
    \includegraphics[width = 0.9\textwidth]{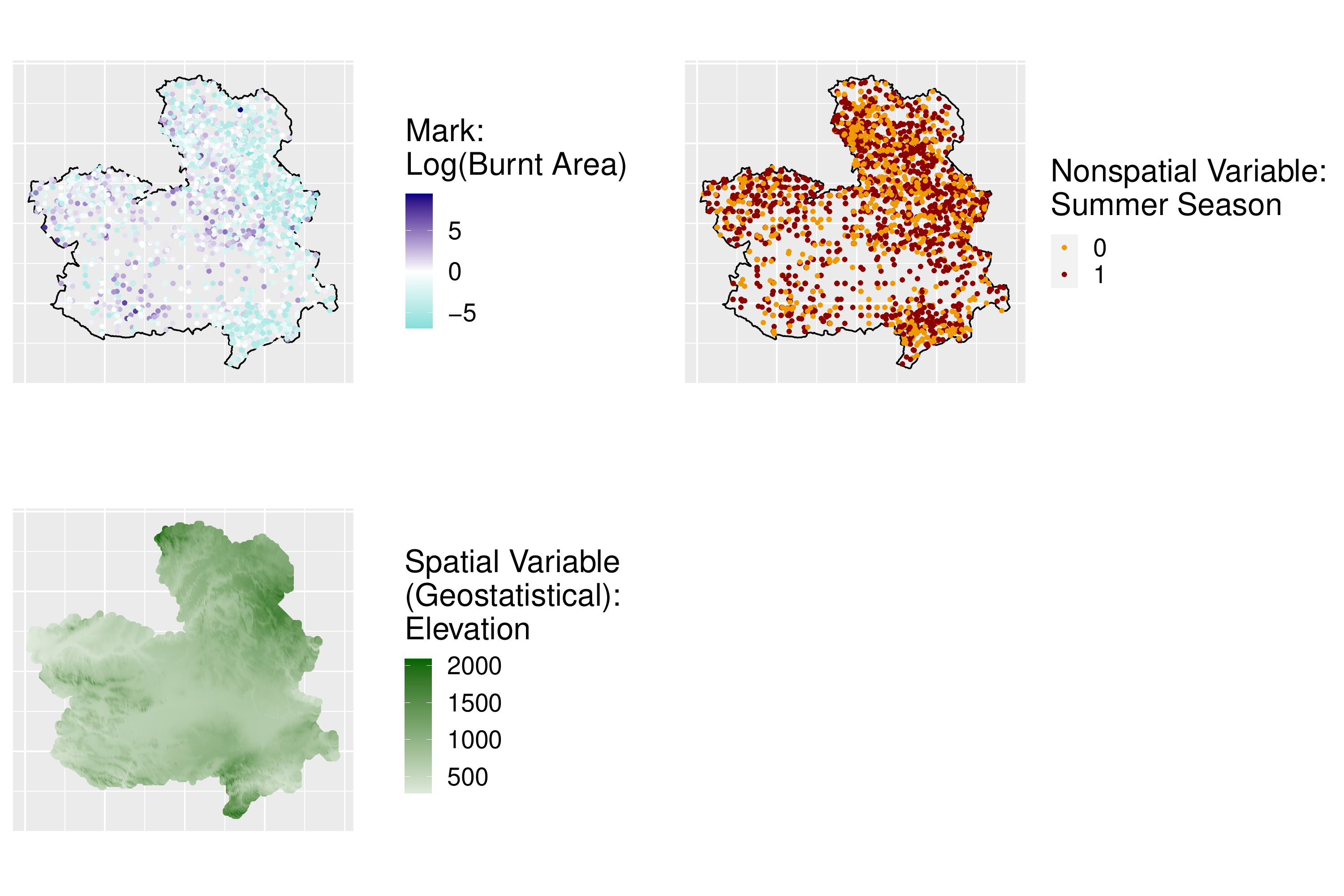}
    
    \caption{\normalsize{Forest Fire Data in Castilla-La Mancha, Spain}}
    \label{fig:firedata}
\end{figure}

In the second motivating example of forest fires, potential spatial variables of interest include elevation and terrain; potential nonspatial variables of interest include the time of year that the fire occurred and the cause of the fire (intentional or not). We are interested in determining the potential effect of these variables on the location of fires and the amount of burned area at those locations (the mark).  Data from this application for the Castilla-La Mancha region of Spain is shown in Figure \ref{fig:firedata}. We note that our model can incorporate multiple types of marks which is illustrated by our applications (categorical for use of force, continuous for forest fires) and multiple types of spatial variables (areal for use of force, geostatistical for forest fire data).

There is considerable attention and statistical methodology for point process models involving spatial variables, such as the log Gaussian Cox process (LGCP) \citep[cf.][]{moller1998log, moller2003statistical}. These LGCP studies have included a wide variety of applications including crime, disease cases, agriculture, and seismic activity \citep{shirota2017space, li2012log, brix2001space, siino2018joint}. There is also some work in marked point processes incorporating spatial dependence in the marking process, where the mean and the variance of the mark is dependent on the spatial intensity function \cite{myllymaki2009conditionally}. However, there has been less extensive work on incorporating both spatial and nonspatial variables into marked point process models. Furthermore, it is important to have a computationally tractable inferential procedure for this model, and we would like the resulting inferences to be easy to interpret.

Here we develop a novel two-stage log Gaussian Cox process (LGCP) model for a marked point process using a Bayesian framework. 
Our model utilizes both event/individual-level (nonspatial) and neighborhood-level (spatial) characteristics. We use this model to determine the impact of spatial variables on the locations of events and the relationship between both kinds of variables and the mark of the point process. We note that some studies may refer to all data available only at point process locations as marks. However, we find that for clarity in our framework developed here, there is value in distinguishing between dependent and independent variables in the marking process. We refer to independent variables as nonspatial variables, and dependent variables as marks. 
We also allow for dependence between the location and mark determination processes. We find that inference for our model is reliable and that our approach provides an intuitive interpretation of the impact of spatial and event/individual-level on marked point processes.

We note that our model provides advantages over existing methods, especially when there is dependence between the location-determination and mark-determination stages of the process. There are multiple ways for dependence to be characterized in marked point processes, including between marks and points, points and the covariates, between marks and the covariates, and between all three \citep[cf.][]{dvovrak2020nonparametric}. Our framework allows users to capture dependence between many of these dependence structures in each stage of the model and between the two stages. We present possibilities for the location-determination stage (NHPP and LGCP) and the mark-determination stage (logistic and linear regression) but there are a vast number of possible models that could be chosen for each stage, especially when the stages are independent. 
Under independence of the location and mark determination processes, their analysis can also be treated independently by classical point process methods for locations and by spatial regression methods for marks. Our regression framework, which allows for simple incorporation of dependent Gaussian processes in both stages of the model, creates a method for further evaluating dependence between stages after controlling for spatial variables and other dependence structures. It is difficult to perform this extent of diagnosis and evaluation of dependence between point locations, marks, and covariates using existing methods.

Previous approaches to incorporating nonspatial variables into a point process framework have typically been in the context of public health data. \cite{liang2008analysis} addresses spatial misalignment between spatial and nonspatial variables with an application to disease cases. Specifically, the authors characterize spatial variation of disease occurrence to identify areas with elevated risk. Their bivariate Gaussian process approach, which we call the bivariate mark model in this paper, breaks new ground by specifying an approach for handling nonspatial covariates in point process models as well as allowing for dependence between marks. Our simulated examples suggest that their approach may pose some parameter identifiability and interpretability issues. \cite{best2000spatial} studies individual-level outcomes of disease cases. The authors utilize individual-level covariate information from patients, such as sex and home conditions like dampness or parental smoking, as well as a spatial random field to account for unattributed spatially varying risk. They also incorporate the level of nitrogen dioxide, which is considered to be spatially varying. The model incorporates both individual-level and spatial variables in order to compare baseline and relative risk, under given sets of conditions. As in \cite{liang2008analysis}, the model measures both location and attribute-specific risk intensity simultaneously, which makes it difficult to describe the spatial distribution of the outcome without choosing particular individual-level attributes.

\cite{quick2015bayesian} uses both spatial and nonspatial information to study mortality rates in an LGCP framework, similar to the model proposed by \cite{liang2008analysis}. However, instead of treating nonspatial point-level information as covariates, they treat all nonspatial variables (race, sex, education, and cause of death) as marks through a cross-tabulation of all possible combinations of categorical nonspatial variables to create 24 categorical marks, each of which is called a group. The intensity is then estimated for each of these 24 combinations/groups. The method is applied to North Carolina mortality data, where the authors analyze the relationship between variables such as education and cause of death, where both are included in this cross-tabulation of marks. This is a rich model but parameter interpretation can pose some challenges, as it depends on groups that are defined by the specific combination of categorical marks, and it is difficult to incorporate continuous explanatory variables. 
As the number of nonspatial variables in the model increases, the number of groups would continue to grow, leading to interpretability and computational challenges.  

\cite{pinto2015point} studies cerebrovascular diseases (CVD) and the nonspatial/ individual-level variables that may play a role in the intensity of cases, namely age at death, education level, gender, and marital status. These nonspatial covariates are incorporated into the intensity function in a similar way to the model proposed by \cite{liang2008analysis} but \cite{pinto2015point} allows for additional spatial variation in the effect of the nonspatial covariates. The model includes three Gaussian processes: one for nonspatial variables, another for the interaction between spatial and nonspatial variables, and a third for an unmeasured spatial random effect. This is a very flexible model formulation. However, we are more interested in the effect of individual-level/nonspatial variables on the mark of the point process than on how the effect of these variables change over space. The model presented by \cite{pinto2015point} is not extended to marked point processes. Furthermore, their flexible model results in large computational challenges, which the authors address by discretizing the analysis into sub-regions.

\noindent We summarize the main contributions of this paper below.
\begin{itemize}[nosep]
        \item We introduce a novel two-stage log Gaussian Cox process model that assigns locations based on spatial predictors in the first stage and marks based on nonspatial and/or spatial variables in the second stage. We test the inclusion of dependence structures both within and between stages of the model. The approach of \cite{liang2008analysis} relies on a categorical mark (or multitype) point process framework. Our two-stage marked point process model allows for both multiple types of spatial variables, including areal and geostatistical, as well as multiple types of marks, including categorical and continuous.  
        \item  Current models \citep[cf.][]{liang2008analysis, diggle2010estimating, pinto2015point} that incorporate nonspatial variables by and large assume that they are available everywhere in the spatial domain by including them in the spatial intensity function. Such an assumption imposes some challenges conceptually as well as in practice. In particular, it becomes challenging to think of a straightforward approach to simulating from such models. We describe how to simulate point process data where the intensity is a function of both spatial and nonspatial variables; this is likely to be broadly useful for practitioners and perhaps also useful in specific applications such as synthetic data creation for preserving privacy \citep[cf.][]{quick2015bayesian}.
        \item We use simulation to understand the interpretability of inference based on our method, and compare it to the bivariate mark modeling approach \citep{liang2008analysis}. We illustrate that incorporating nonspatial variables into a spatial intensity function imposes challenges for simulation and inference. Our approach avoids using nonspatial variables in this way by separating the location and mark determination stages of the model, where nonspatial variables are only used to determine the mark of the point process. Furthermore, we study how the availability of replicates may impact parameter inference.
\end{itemize}

The paper is organized as follows. In Section \ref{sec_2}, ``Marked Point Process Models with Spatial and Nonspatial Predictors'', we summarize the modeling framework for marked point processes generally. We give detailed specifications of our new two-stage model, including an outline of a simulation algorithm. We also present an existing bivariate marked point process model that we will compare to the two-stage model \citep{liang2008analysis}.  In Section \ref{app_sec}, ``Applications to Simulated Data'', we outline a study of both our new two-stage model and the bivariate mark model to simulated data in order to test the ability of these models to recover simulated parameters. Next, we apply these methods, in a variety of specifications, to real data in Section \ref{sec:fire_data}, ``Application to Castilla-La Mancha Fire Data.'' In the discussion section, Section \ref{sec_disc}, we summarize our results, the benefits and limitations of our model and potential for future work. We relegate to the appendix some of the details of computation and a related model, the bivariate mark model \citep{liang2008analysis}.

%%%%%%%%%%%%%%%%%%%%%%%%%%%%%%%%%%%%%%%%%%%%
%%%%%%%%%%%%%%%%%%%%%%%%%%%%%%%%%%%%%%%%%%%%
%%%%%%%%%%%%%%%%%%%%%%%%%%%%%%%%%%%%%%%%%%%%
\section{Marked Point Process Models with Spatial and Nonspatial Predictors}
\label{sec_2}

We describe the general setup for log Gaussian Cox Process (LGCP) models in order to motivate two models: our new two-stage marked point process model and the bivariate mark model developed by \cite{liang2008analysis}. Point processes are defined by a set of n locations $s_1,...,s_n$ in a spatial region/window of interest, $W$. In our case, these $n$ locations are police use of force incidents in Dallas, Texas or forest fires in the Castilla-La Mancha region of Spain. We assume the $n$ events come from a spatial point process with intensity $\lambda_{\bm{\theta}}(s)$, $s \in W$, where $\bm{\theta}$ consists of unknown parameters such as regression coefficients and covariance function parameters. 
 The intensity may be determined by a vector of spatial covariates at location $s$, which we notate as $\bm{z(s)}$. The spatial variables for our study of police use of force in Dallas are collected at the census tract level and points are assigned the census estimate of the census tract where they are located. Spatial variables for the forest fire data are geostatistical, and are provided on a fine grid over the region. Locations of the point process are assigned values for spatial variables of the nearest grid location. The intensity may also be impacted by a vector of individual/event-level or nonspatial covariates at that event, which are denoted $\bm{\nu}$ \citep[cf.][]{liang2008analysis}.

 For a log Gaussian Cox process \citep[cf.][]{moller2003statistical}, the intensity function $\lambda_{\bm{\theta}}(s)$ may be defined as follows: 
 $\log(\lambda_{\bm{\theta}}(s)) = \bm{z(s)}'\bm{\beta} +\bm{\nu}'\bm{\alpha} + \omega(s)$, where $\bm{\theta}$ consists of $\bm{\beta}$ and $\bm{\alpha}$, which are vectors of unknown spatial and nonspatial regression coefficients respectively. The spatial intensity function is typically defined with spatial variables ($z(s)$) but has incorporated nonspatial variables ($\nu$) in recent work \citep[cf.][]{liang2008analysis}. The vector $\bm{\theta}$ also consists of parameters of $(\omega(s), s \in W)$, which is a zero-mean  Gaussian process. We define the Gaussian process via a positive definite covariance function. Later we provide details about the form of the covariance function we use for each model. For estimation of the Gaussian processes in this paper, we rely on the predictive process approach, as outlined in Section \ref{pred_proc_sec}.

The resulting likelihood function is given below, where the parameter vector $\bm{\theta}$ typically consists of regression coefficients, such as  $\bm{\beta}$ and $\bm{\alpha}$ as described above, and parameters of the covariance function. Details of the estimation of the integral are given in \ref{app_sec:integ_methods}.

\begin{equation}
     \mathcal{L}(\bm{\theta}; s_1,...,s_n, s\in W) \propto  \exp\left(-\int_{W} \lambda_{\bm{\theta}}(s)ds\right) \times \prod_{i=1}^n \lambda_{\bm{\theta}}(s_{i}) 
    \label{lgcp_lik}
\end{equation}

We denote the mark at each point process location as $Y(s_i)$, which could be continuous or categorical. When marks are categorical, such as type of force used by police, we call these point processes multitype point processes. We also consider a continuous mark in the case of forest fire data, where we model the amount of burned area in each fire. Existing approaches are often specified specifically for categorical (multitype) point processes \citep[cf.][]{waagepetersen2016analysis, moller1998log, brix2001space, liang2008analysis}.  We develop a framework that is flexible and can easily accommodate multiple kinds of marks, including continuous and categorical (multitype) marks.

We use a Bayesian inferential approach for these models with computation done by Markov Chain Monte Carlo (MCMC). Both the two-stage and the bivariate mark models are fit using the NIMBLE R package, which allows for flexible Bayesian computation \citep{nimble}. We assess convergence of our MCMC-based approximations through the use of trace plots and effective sample sizes, with the latter based on rigorous Markov Chain Monte Carlo standard error estimates \citep{batchmeans_package, mcmcse_package}. Detailed prior distributions are given in the description of the two-stage and bivariate mark models. Our MCMC approach uses Metropolis-Hastings adaptive random-walk samplers with univariate normal proposal distributions for the regression coefficients and multivariate normal proposals for the Gaussian processes \citep{nimble}.

%%%%%%%%%%%%%%%%%%%%%%%%%%%%%%%%%%%%%%%%%%%%
%%%%%%%%%%%%%%%%%%%%%%%%%%%%%%%%%%%%%%%%%%%%
%%%%%%%%%%%%%%%%%%%%%%%%%%%%%%%%%%%%%%%%%%%%

\subsection{Two-Stage Marked Point Process Model}
\label{sec:two_stg_desc}
The two-stage model we develop here specifies a first stage model that determines the locations of the point process events and a second stage which determines the marks at those locations.
 Two-stage models, sometimes referred to as hurdle models, have been developed in the context of areal data \citep[cf.][]{ver2007space, michaud2014estimating} %, neelon2013spatial} 
 as well as point-referenced data \citep[cf.][]{recta2012two}.

In our model, the nonspatial covariates are not used to determine the locations of the events, but rather serve as part of the mark-determination stage. Nonspatial covariates may not be available across the spatial domain but are typically only observed where a point process event occurs, which motivates this model structure. Nonspatial variables may be associated with an event (such as arrest/hospitalization at an incident) or with an individual involved in an event (officer/citizen race/gender). The first stage of this model consists of an LGCP based on spatial variables and a Gaussian process, where the intensity takes the following form: $\log(\lambda_{\bm{\theta}}(s)) = \bm{z(s)'\beta} + \omega(s)$. For the Gaussian process, $\{\omega(s), s \in W\}$, we assume a univariate exponential covariance function so that for any $s_i, s_j \in W$ the covariance is defined such that $\text{Cov}(\omega(s_i), \omega(s_j))  = \sigma^2 \exp\left(\frac{-|s_i-s_j|}{\phi}\right) $, with parameters $\sigma^2, \phi > 0$.

The second stage of the model, in the case of a binary categorical mark, consists of logistic regression to determine the mark, given the locations. The second stage of the model could include any combination of nonspatial and spatial variables. In addition to logistic regression, the second stage of our two-stage model allows for multinomial models, linear regression, and other model types to accommodate marks of many forms. The logistic regression for a binary mark takes the following form, where $Y(s_i)$ is the mark of location $s_i$ (either 0 or 1): 

\vspace{-6pt}

$$ \text{logit}(P(Y(s) = 1)) = \bm{z(s)}'\bm{\gamma} +\bm{\nu}'\bm{\alpha}.$$

Depending on the particulars of the application, spatial predictors can be included or excluded from the second stage of the model. This two-stage model may be more appropriate than alternatives when the relationship between covariates (especially nonspatial covariates) and marks are of interest, rather than the dependence between types of points. Dependence between types of points may be better captured through a bivariate Gaussian process approach as described in \cite{liang2008analysis}.

Our two-stage marked point process framework can include different structures for dependence both within and between the stages of the model, which are shown in Table \ref{tab:model_type}. In the first stage of the model, we investigate the cases where there is no additional spatial structure outside of spatial covariates (Model 1), spatial structure through a univariate Gaussian process (Models 2 and 3), and spatial structure that is dependent on the marking process through a bivariate Gaussian process (Model 4). In the second stage, we study the cases where there is no spatial structure in the second stage of the model (Models 1 and 2), as well as a univariate GP (Model 3) and bivariate GP (Model 4) as in the first stage.

 In the most general and flexible formulation of our model, we allow for spatial dependence between the two stages of our model, or Model 4 in Table \ref{tab:model_type}. We assume a Gaussian process for the error in both stages of the model, and then incorporate a dependence structure between the two Gaussian processes. Following \cite{liang2008analysis}, the Gaussian processes $\{\omega_1(s), s \in W\}$ and $\{\omega_2(s), s \in W \}$ are defined such that $(\omega_1(s), \omega_2(s))' \sim MVGP(\bm{0}, \Gamma(\cdot, \cdot, \theta))$ where the cross-covariance matrix $\Gamma(s_i, s_j, \theta)$ is of dimension $2n \times 2n$, where $n$ is the number of points. The cross-covariance matrix consists of both a univariate exponential correlation function, as defined above where $g(s_i, s_j, \phi) = \exp\left(\frac{-|s_i-s_j|}{\phi}\right)$, and the matrix $\Lambda$, as defined below. The full cross-covariance function is $\Gamma(s_i, s_j, \theta) = g(s_i, s_j, \phi)\Lambda$ \citep[cf.][]{liang2008analysis}. If $|\rho|$ is close to 1, this would indicate strong dependence, and close to 0 would indicate weak dependence.

$$
\Lambda = 
\begin{bmatrix}
\sigma_{1}^2 & \rho\sigma_1\sigma_2\\
\rho\sigma_1\sigma_2 & \sigma_{2}^2\\
\end{bmatrix}
\quad
$$

\begin{table}
\centering
\begin{tabular}{|l|ll|r}
 \hline
 Model Number & First Stage &  Second Stage  \\ 
  \hline
Model 1 & NHPP & $\epsilon \sim N(0, \sigma^2)$ \\
Model 2 & Univariate Gaussian Process & $\epsilon \sim N(0, \sigma^2)$ \\
Model 3 & Univariate Gaussian Process & Univariate Gaussian Process \\
Model 4 & \multicolumn{2}{c|}{Bivariate Gaussian Process}  \\
   \hline
\end{tabular}
\caption{\label{tab:model_type} Two-Stage Models Considered. In the case of linear regression (continuous mark), we consider standard normal error in the second stage of the model, as shown in the table. In the case of logistic regression (categorical mark), we consider no error term in the second stage of the model.} %\\
\end{table}

This formulation leads to the following two-stage model, described in Equation \ref{eq:dep_model}, with the first stage determining the locations and the second stage determining the mark of the point process.  The dependence introduced by the Gaussian processes allows for dependence in the spatial structure that determines both where points occur and the mark at those point locations.

\begin{comment}
\begin{equation}
    \begin{split}
        \text{ } & \log(\lambda_{\bm{\theta}}(s))  = \bm{z(s)'\beta} + \omega_1(s) \\
        
        \text{ } & \text{logit}(P(Y(s_i) = 1)) = \bm{z(s)}'\bm{\gamma} +\bm{\nu}'\bm{\alpha} + \omega_2(s) \\
    \end{split}
    \label{eq:dep_model}
\end{equation}
\end{comment}

\begin{equation}
        \begin{split}
        &\log(\lambda_{\bm{\theta}}(s))  = \bm{z(s)'\beta} + \omega_1(s) \\
        &\text{logit}(P(Y(s_i) = 1)) = \bm{z(s)}'\bm{\gamma} +\bm{\nu}'\bm{\alpha} + \omega_2(s) 
        \end{split}
        \label{eq:dep_model}
\end{equation}

The likelihood for the two-stage model, including the logistic regression portion, is included below, where $p_{\bm{\theta}}(s_i) = P(Y(s_i) = 1)$ and $\bm{\theta} = (\bm{\alpha}, \bm{\beta}, \bm{\gamma}, \sigma)$.

\begin{equation}
    \begin{split}
    \mathcal{L}(\bm{\theta}; s_1,...,s_n, s\in W)  \propto  \underbrace{\exp(-\int_{W} \lambda_{\bm{\theta}}(s)ds) \times
    \prod_{s_{i}} \lambda_{\bm{\theta}}(s_{i})}_{\text{Location Determination Stage}} \times \\
    \underbrace{\prod_{s_i} p_{\bm{\theta}}(s_i)^{Y(s_i)}(1-p_{\bm{\theta}}(s_i))^{1-Y(s_i)}}_{\text{Mark Determination Stage}}
    \end{split}
    \label{eq:two_stage_mod}
\end{equation}

In our Bayesian hierarchical framework, we assume Normal(0,100) priors for all of the regression coefficients ($\bm{\alpha}, \bm{\beta}, \bm{\gamma}$). We use an Inverse-Gamma($\alpha = 2$, $\beta = 0.5$) prior for $\sigma$, as in \cite{liang2008analysis}. As in other studies \citep[cf.][]{liang2008analysis}, we fix $\phi$ due to identifiability issues. We want the covariance function not involving $\sigma$, $\exp (-|s_i-s_j|/\phi)$, to be small when points are far apart and large when points are close together. We find the value of $\phi$ so that the 95$^{th}$ percentile of distances would have a a correlation of 0.05, and the value of $\phi$ so that the 5$^{th}$ percentile of distances would have a a correlation of 0.95. We fix $\phi$ at the average of these two values.

\begin{algorithm}[ht]
  \caption{Two-Stage Marked Point Process Simulation}
  \label{twostage_algorithm}
  \begin{algorithmic}[1]
    \State Define $n$ locations, $x_1,\dots x_n \in W$.$^\dag$ Simulate a Gaussian process, $\omega(x_1), \dots, \omega(x_n)$ with parameters $\sigma^2, \phi$.
    \State Evaluate the intensity $\lambda_{\bm{\theta}}(s) = \bm{z(s)}'\bm{\beta} + \omega(s)$ at $x_1,\dots x_n$ $\hspace{-5pt}^{\dag\dag}$ and \textbf{compute the maximum intensity}, $\lambda_{\max} = \max(\lambda_{\bm{\theta}}(x_1) \dots, \lambda_{\bm{\theta}}(x_n))$.
    \State \textbf{Simulate a homogeneous Poisson Process} with parameter $\lambda_{\max}\times a$ on $W$, where $a$ is the area of $W$.
    \State Affiliate each simulated point with one of $x_1,\dots,x_n$.\hspace{0pt}$^\ddag$ This is to assign a value of $\omega(s)$ to each simulated point of the homogeneous Poisson process.
    \State Calculate the probability of keeping the point at location s, $\lambda_{\bm{\theta}}(s)/\max(\lambda_{\bm{\theta}}(s))$. \textbf{Thin the points} by keeping each point with this probability, otherwise remove.
    %%%%%
    \State Simulate nonspatial variables according to their distribution and assign values to each point from the thinned homogeneous Poisson process.
    \State \textbf{Assign the location s to mark 1} with probability, $p_{\bm{\theta}}(s)$, where $\text{logit}(p_{\bm{\theta}}(s))= \bm{z(s)}'\bm{\gamma} +\bm{\nu}'\bm{\alpha}$. Otherwise, assign to mark 0.
    
    \vspace{12pt}
    \hspace{-40pt} $\dag$: In our approach, we found it helpful to generate $x_1,\dots x_n$ based on a grid of equally spaced points in the observation window, $W$.
    
    \hspace{-40pt} $\dag\dag$: When computing the maximum data using spatial variables available on the areal level, at least one point per areal unit is needed. We use a number of points per unit proportional to the area of the areal unit.
    
    \hspace{-40pt} $\ddag$: Options for affiliating each simulated point with $x_1,\dots,x_n$ include affiliation with the closest point or predictive process transformations.
  \end{algorithmic}
\end{algorithm}

We fit our model to data simulated from the two-stage model via Algorithm \ref{twostage_algorithm}. 
In order to simulate from this model, we assume a spatial window of interest with area $a$ (in the simulated example, it is the Dallas city boundary). For our simulated example, we also assume that $W$ has been broken into areal units and we have spatial variables affiliated with those areal units. We use the Census data available to us from census tracts, but these variables may also be simulated. Spatial data could also be geostatistical, as with the elevation data in the forest fire dataset. We now describe in Algorithm \ref{twostage_algorithm} how to simulate from a two-stage marked point process with parameters $\alpha_1, \dots, \alpha_p$, the regression coefficients for the nonspatial variables, $\beta_1,\dots, \beta_k $, the coefficients for the spatial variables that are used in the first stage, $\gamma_1, \dots, \gamma_r$, the coefficients for the spatial variables that are used in the second stage, and covariance parameter $\sigma >0$. We simulate nonspatial variables from Beta and Bernoulli distributions, but we find that the two-stage model is not sensitive to choice of distribution when other distributions are tested. 

\begin{figure}
\captionsetup[subfigure]{justification=justified, singlelinecheck=false}
\subfloat[Calculate $\boldsymbol{\lambda_{\max}}$: \\          $\log(\lambda_{\bm{\theta}}(s)) = \bm{z(s)}'\bm{\beta} + \omega(s)$]{\label{fig:hurd-sub-first}%
        \includegraphics[height=1.2in]{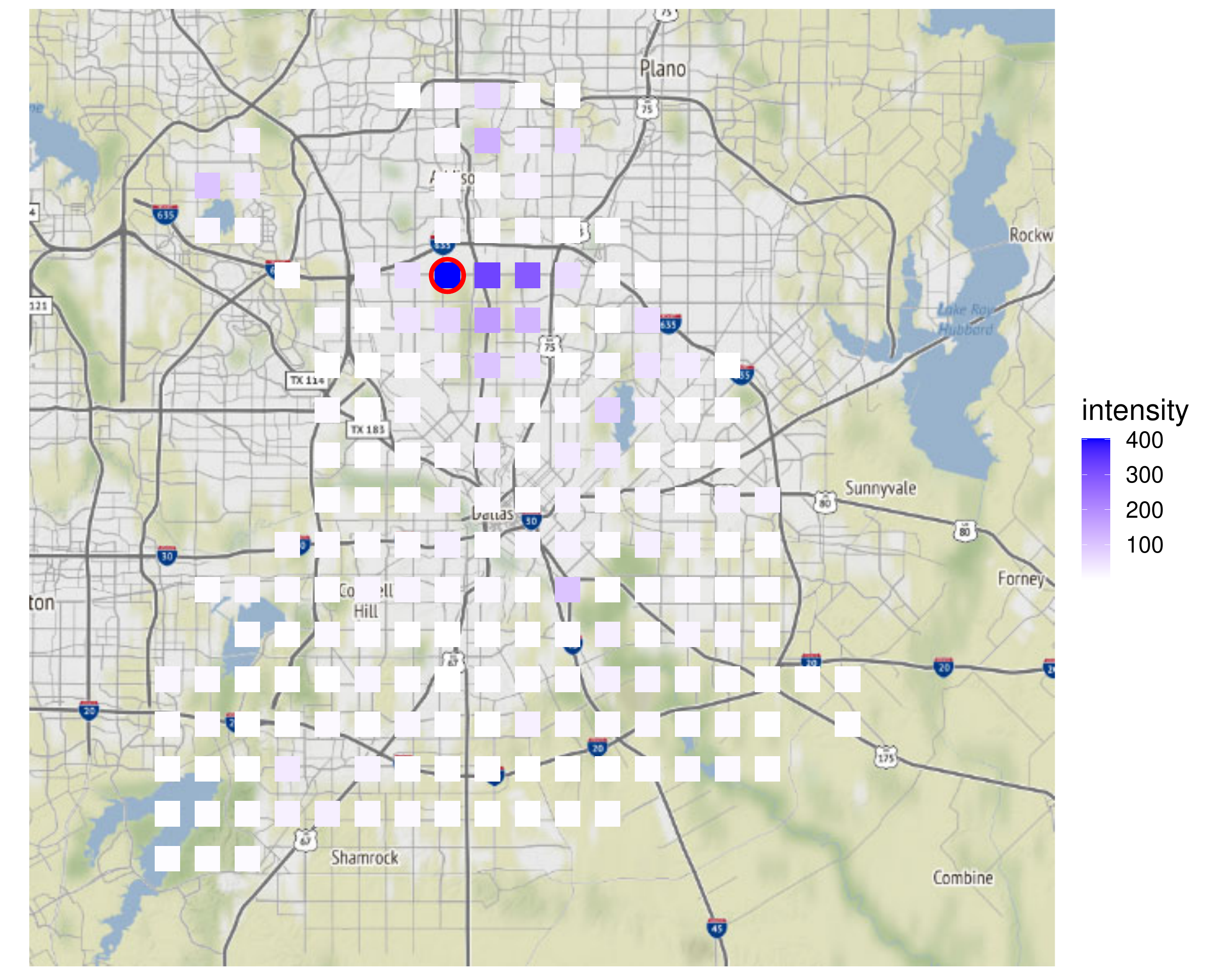}
}%
%\hfill
\hspace{1.8in}
\subfloat[Simulate Homogeneous Poisson \\ Process:
         $n \sim \text{Poisson}(\lambda_{\max}\times |W|)$]{\label{fig:hurd-sub-second}%
    \includegraphics[height=1.3in]{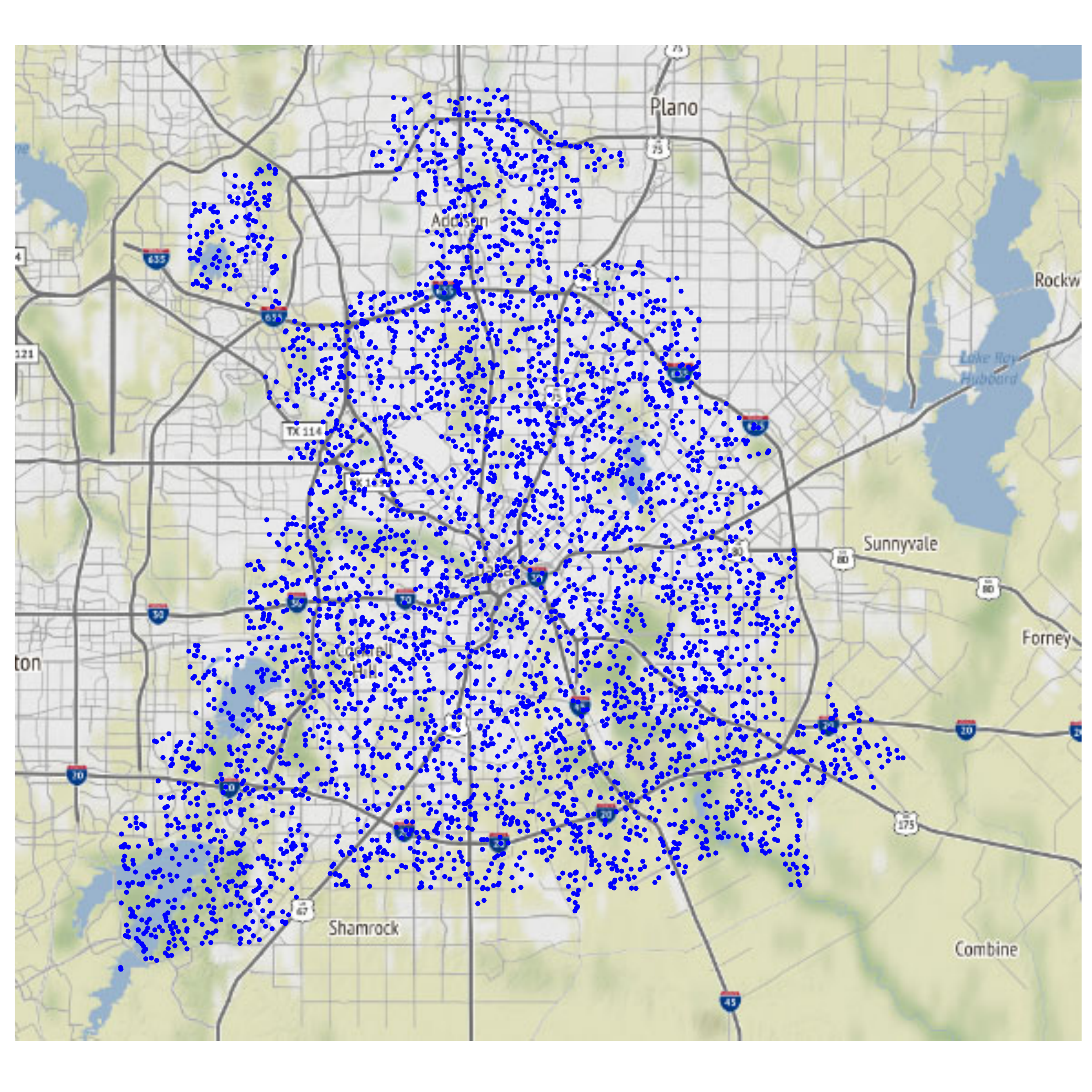}    %
}

\vspace{-10pt}

\subfloat[{Thinning Based on Spatial \\ Variables:
         $P(\text{keep s}) = \lambda_{\bm{\theta}}(s)/\max(\lambda_{\bm{\theta}}(s))$}]{\label{fig:hurd-sub-third}%
         %\centering
        %\hspace{-1in}
        %\centering
        %\hspace{0.2in}
        \includegraphics[height=1.2in]{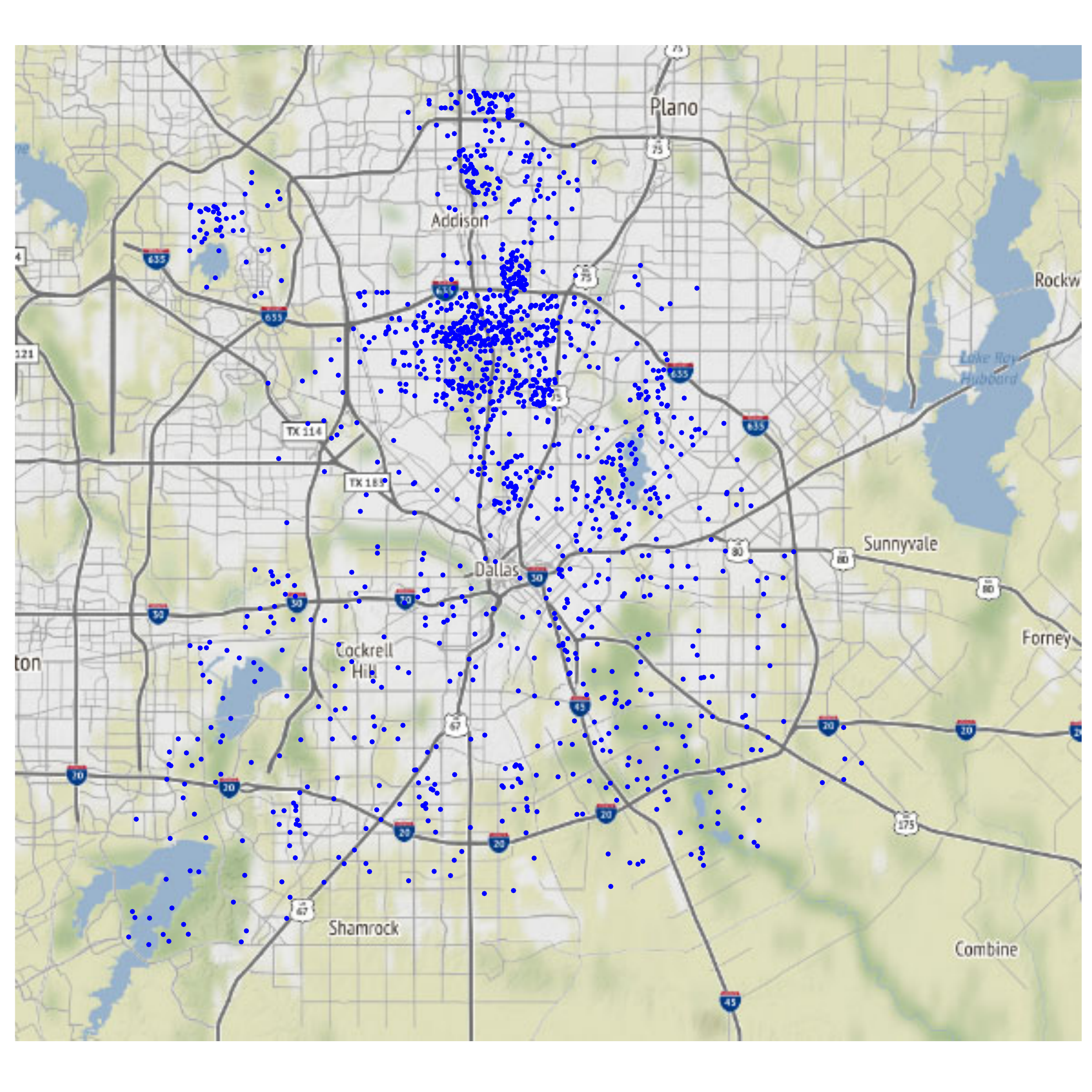} 
}%
%\hspace{1.4in}
%\hfill
\hspace{2.1in}
\subfloat[{Assign Mark:} \\ $\text{logit}(P(Y(s_i) = 1))= \bm{\nu}'\bm{\alpha} $]{\label{fig:hurd-sub-fourth}%
    %\hspace{0.3in}
    \includegraphics[height=1.2in]{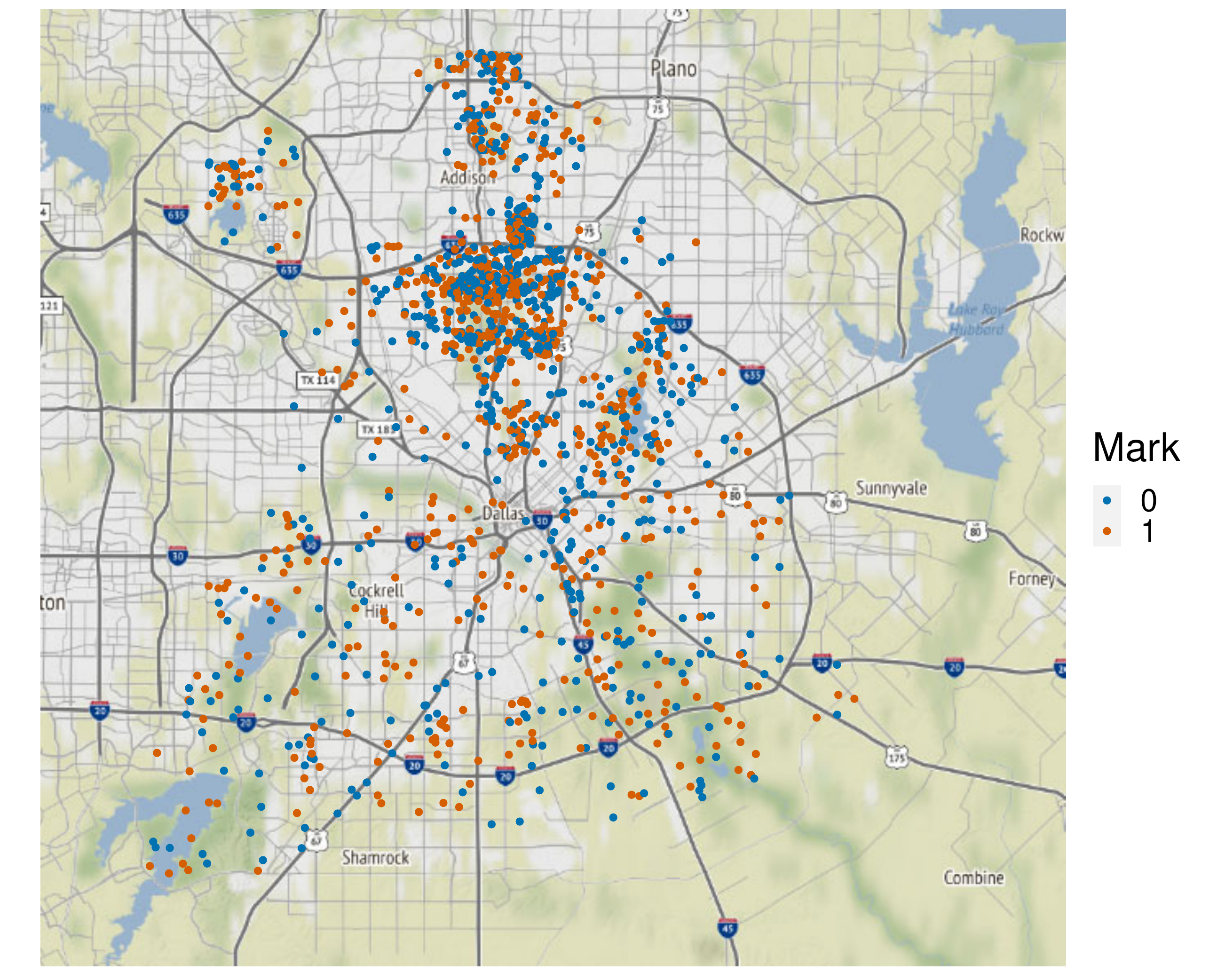}    %
}

\vspace{-10pt}

\caption{Simulation Process for Two-Stage Marked Point Process Model\label{fig:hurd_sim}}
\end{figure}

First, we calculate the maximum intensity across the space ($\lambda_{max}$) in order to simulate a homogeneous Poisson process based on this maximum intensity, as shown in Figure \ref{fig:hurd-sub-first} and \ref{fig:hurd-sub-second}. In order to calculate the maximum intensity, there needs to be at least one point per areal unit, or census tract in our case, in order to ensure that we accurately calculate the maximum intensity. Next, we thin the homogeneous points based on the intensity $\lambda_{\bm{\theta}}(s)$, which in our case only depends on spatial variables, to determine the final locations present in our dataset, which is shown in Figure \ref{fig:hurd-sub-third}. There are no restrictions on the distributional assumptions of the nonspatial variables. For instance, unlike the
 bivariate mark model (Section \ref{sec:bivmodel}), they need not be uniform or bounded. 
After the thinning procedure, both the marks and nonspatial variables are assigned to the remaining locations, as shown in Figure \ref{fig:hurd-sub-fourth}.

\subsection{Bivariate Mark Model}
\label{sec:bivmodel}
%\textcolor{blue}{Check}

A bivariate marked point process model was developed by \cite{liang2008analysis} to study two types of cancer and their relationship to each other and with community-level and individual characteristics. The model includes mark-dependent coefficients for both spatial and nonspatial variables. Dependence between categorical marks is introduced through a dependent Gaussian process structure, where $\{\omega_1(s), s \in W\}$ and $\{\omega_2(s), s \in W\}$ follow the dependence structure shown in Section \ref{sec:two_stg_desc}. Instead of the Gaussian processes being affiliated with each stage as in the two-stage model, the Gaussian processes are affiliated with each level of the mark, or level of force in our simulated example.

In general, the mark-dependent intensity function for marks $k = 1,2$, notated $\lambda_{\bm{\theta}_k}(s, \nu)$, takes the following form. The mark dependent coefficients $\bm{\beta_k}$ correspond to spatial covariates at location $s$, $\bm{z(s)}$. The mark dependent coefficients $\bm{\alpha_k}$ correspond to nonspatial covariates, $\bm{\nu}$. Finally, $\{\omega_k(s), s \in W\}$, $k = 1,2$ are the Gaussian processes affiliated with each level of the mark. The Gaussian processes are estimated in a similar fashion to the Gaussian processes described in our two-stage model, except for the cross-covariance and dependence are across levels of the mark in the bivariate mark model.

$$\log(\lambda_{\bm{\theta}_k}(s, \nu)) = \bm{z(s)}'\bm{\beta_k} +\bm{\nu}'\bm{\alpha_k} + \omega_k(s)$$

The likelihood for the bivariate mark model including both spatial and nonspatial variables and for $k$ levels of the mark is given below, where $\bm{\theta_k} = (\bm{\alpha_k}, \bm{\beta_k}, \sigma_1, \sigma_2, \rho)$:
\begin{equation}
    \begin{split}
     \mathcal{L}(\bm{\theta_k}; s_1,...,s_n, s\in W) \propto & \prod_{k} \exp(-\int_{W}\int_{V} \lambda_{\bm{\theta}_k}(s, v)dvds) \times \prod_{s_{ki}, v_{ki}} \lambda_{\bm{\theta}_k}(s_{ki}, v_{ki}) 
    \end{split}
    \label{bvgp_likelihood}
\end{equation}

The intensity for each mark level $k$, $\lambda_{\bm{\theta}_k}(s,v)$, could be specified with or without the Gaussian process $\omega_k(s)$. The Gaussian process specification allows for the additional flexibility of modeling dependence between the mark levels.

As in the two-stage model, in our Bayesian framework we use Normal(0,100) priors for all of the regression coefficients ($\bm{\alpha_k}, \bm{\beta_k}$ for marks $k = 1,2$). We use an Inverse-Gamma($\alpha = 2$, $\beta = 0.5$) prior for $\sigma_1$ and $\sigma_2$ and a Uniform (-0.999,0.999) prior for $\rho$, as in \cite{liang2008analysis}.

\begin{figure}
\subfloat[{Calculate $\max(\lambda_{\bm{\theta}_k}(s, \nu))$ for $k$ marks:} 
         $\log(\lambda_{\bm{\theta}_k}(s, \nu)) = \bm{z(s)}'\bm{\beta_k} + \bm{\nu}'\bm{\alpha_k}+ \omega_k(s)$]{\label{fig:sub-first}%
         %\centering
        %\hspace{-1in}
        \hspace{-0.05in} \\[-1in]
        \includegraphics[height=1.3in]{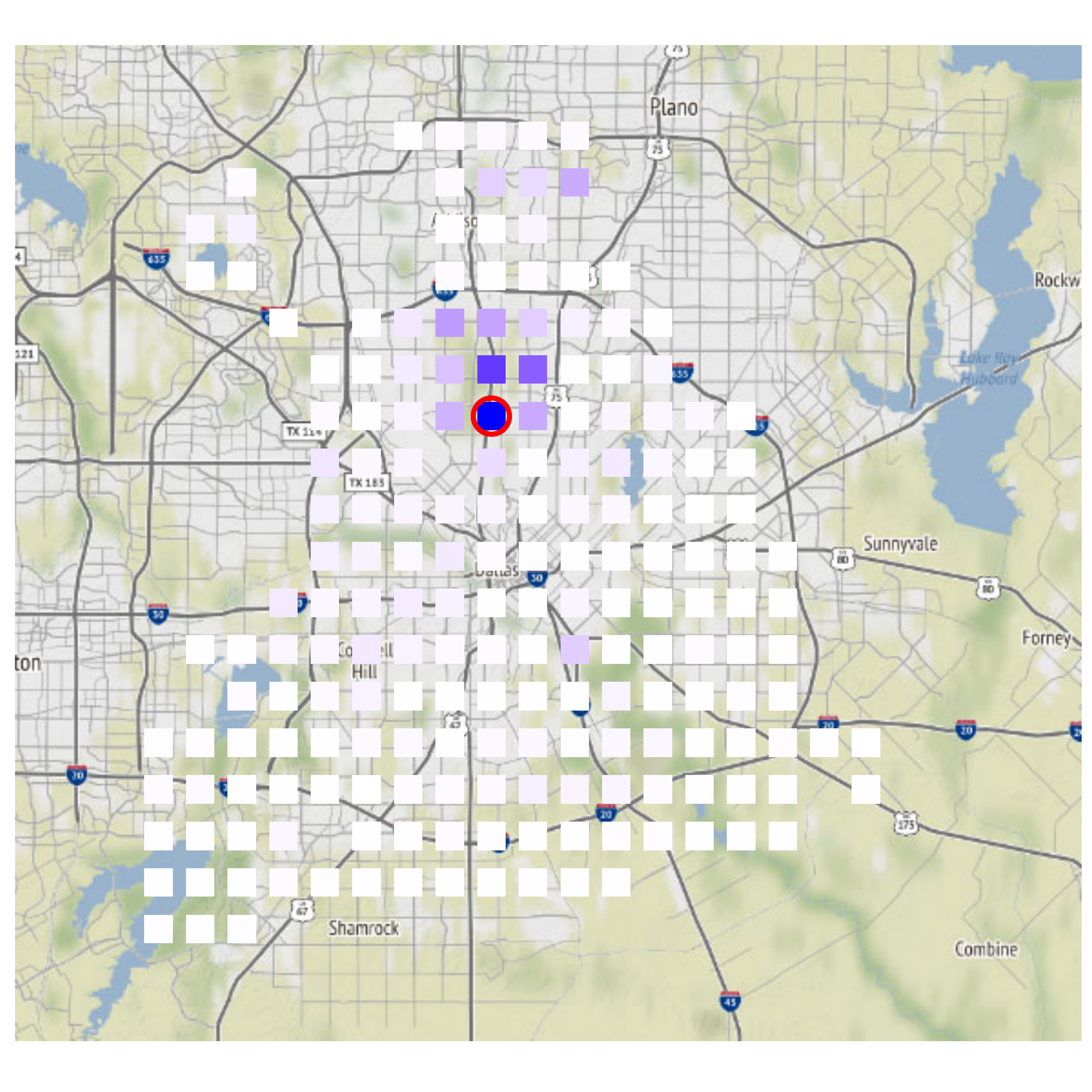}

      %\hspace{1in}
      \hspace{-0.4in}
      \includegraphics[height=1.2in]{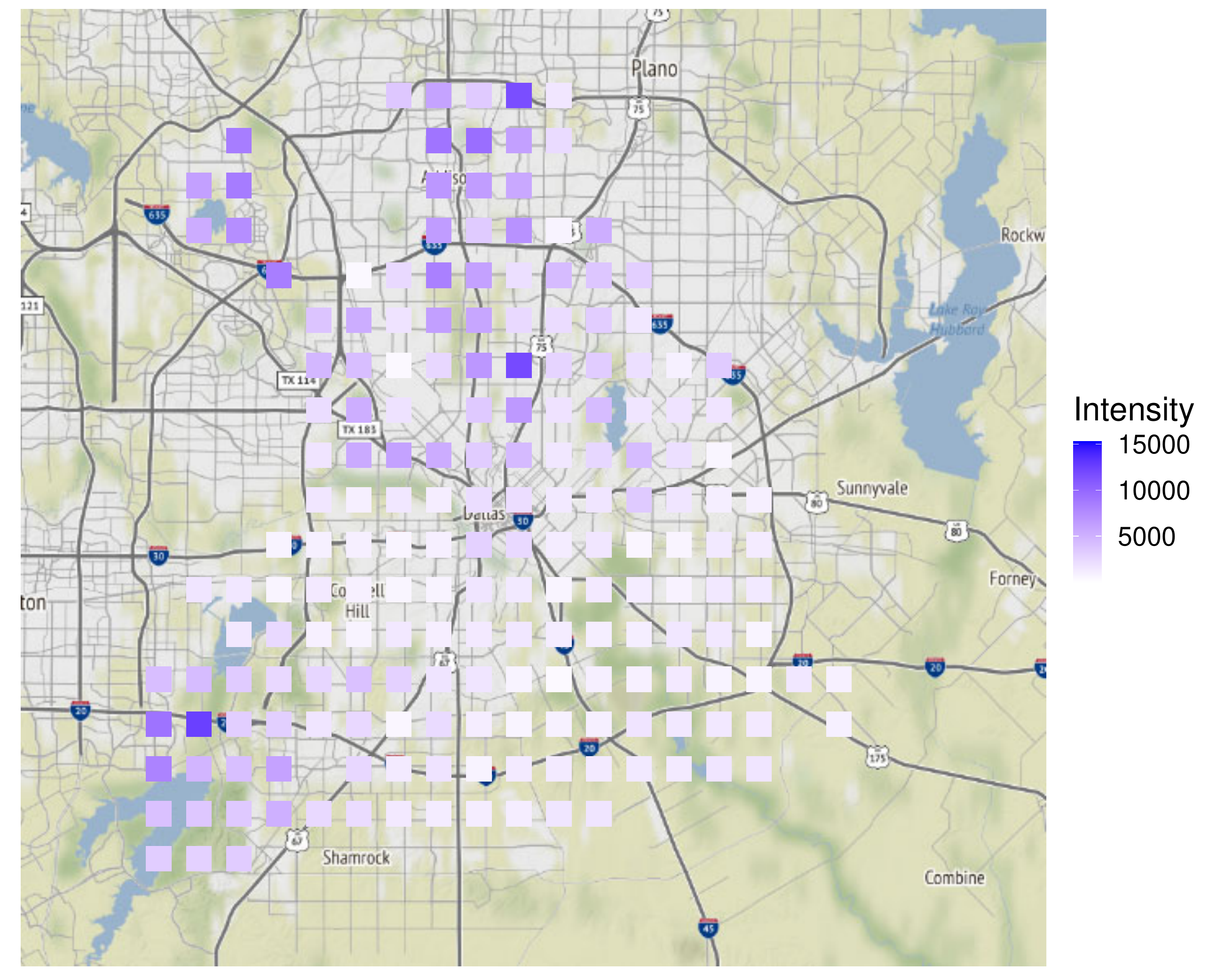} %
}%
\hfill
\subfloat[{Simulate Homogeneous Poisson Processes for $k$ marks: } 
             $n_k \sim \text{Poisson}(\max(\lambda_{\bm{\theta}_k}(s, \nu))*|W|)$]{\label{fig:sub-second}%
  \hspace{-0.1in}\includegraphics[height=1.3in]{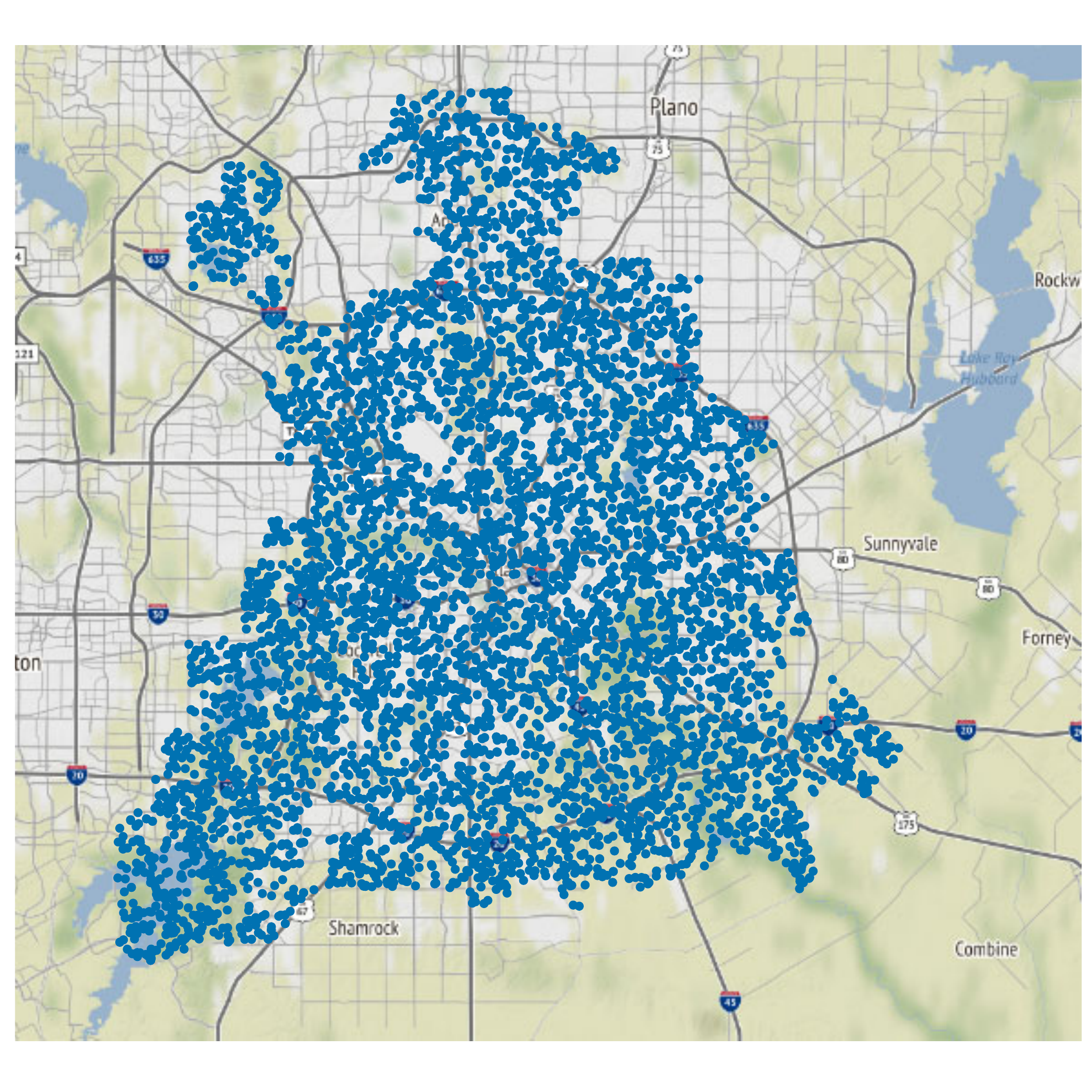}  
      
      \vspace{-1in}
      \hspace{-0.2in}\includegraphics[height=1.2in]{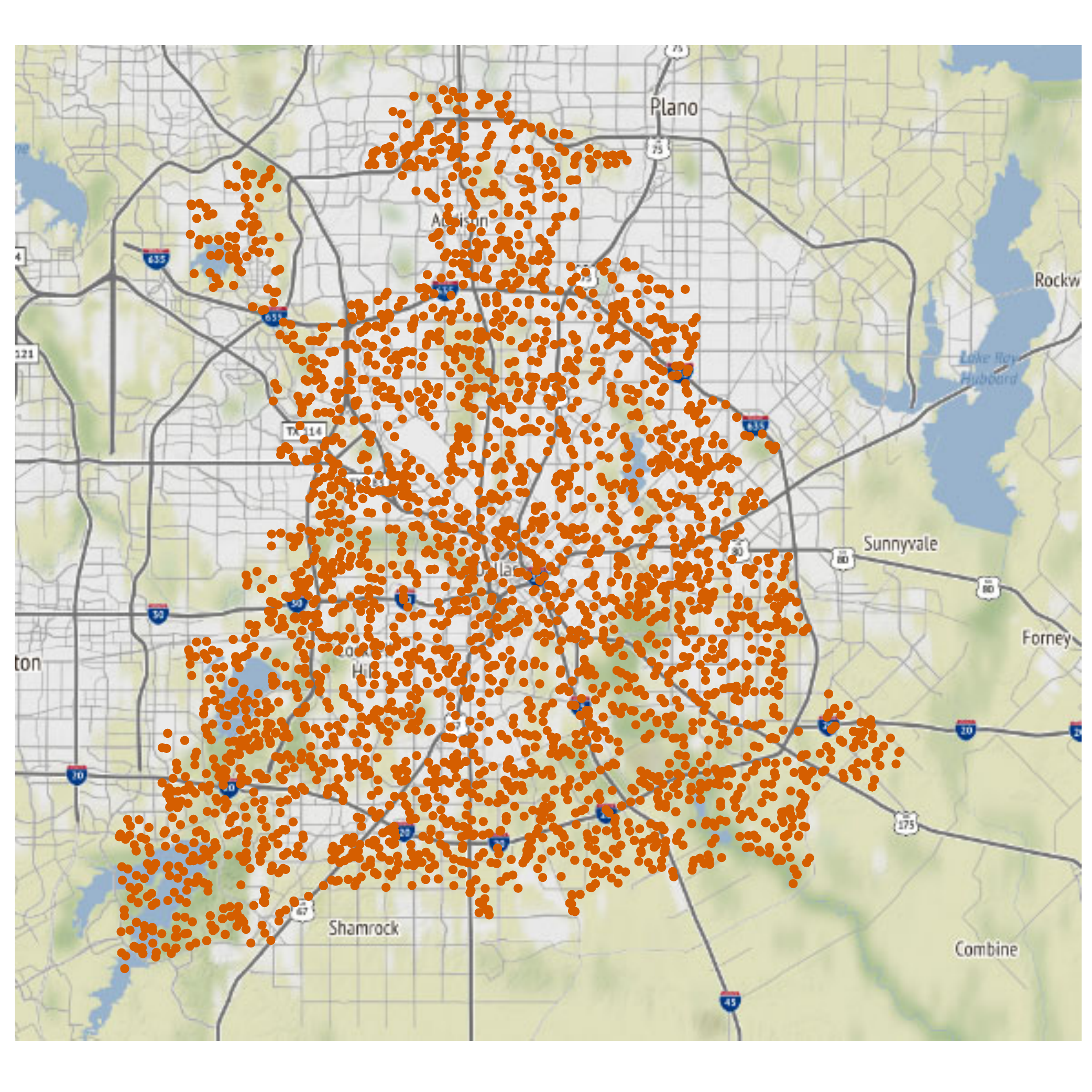}   %
}

\vspace{-10pt}

\hspace{0.3in}\subfloat[{Simulate Nonspatial Variables: }
             Based on \\ uniform  distributions]{\label{fig:sub-third}%
         %\centering
        %\hspace{-1in}
        %\centering
        \hspace{0.2in}
        \includegraphics[height=1.2in]{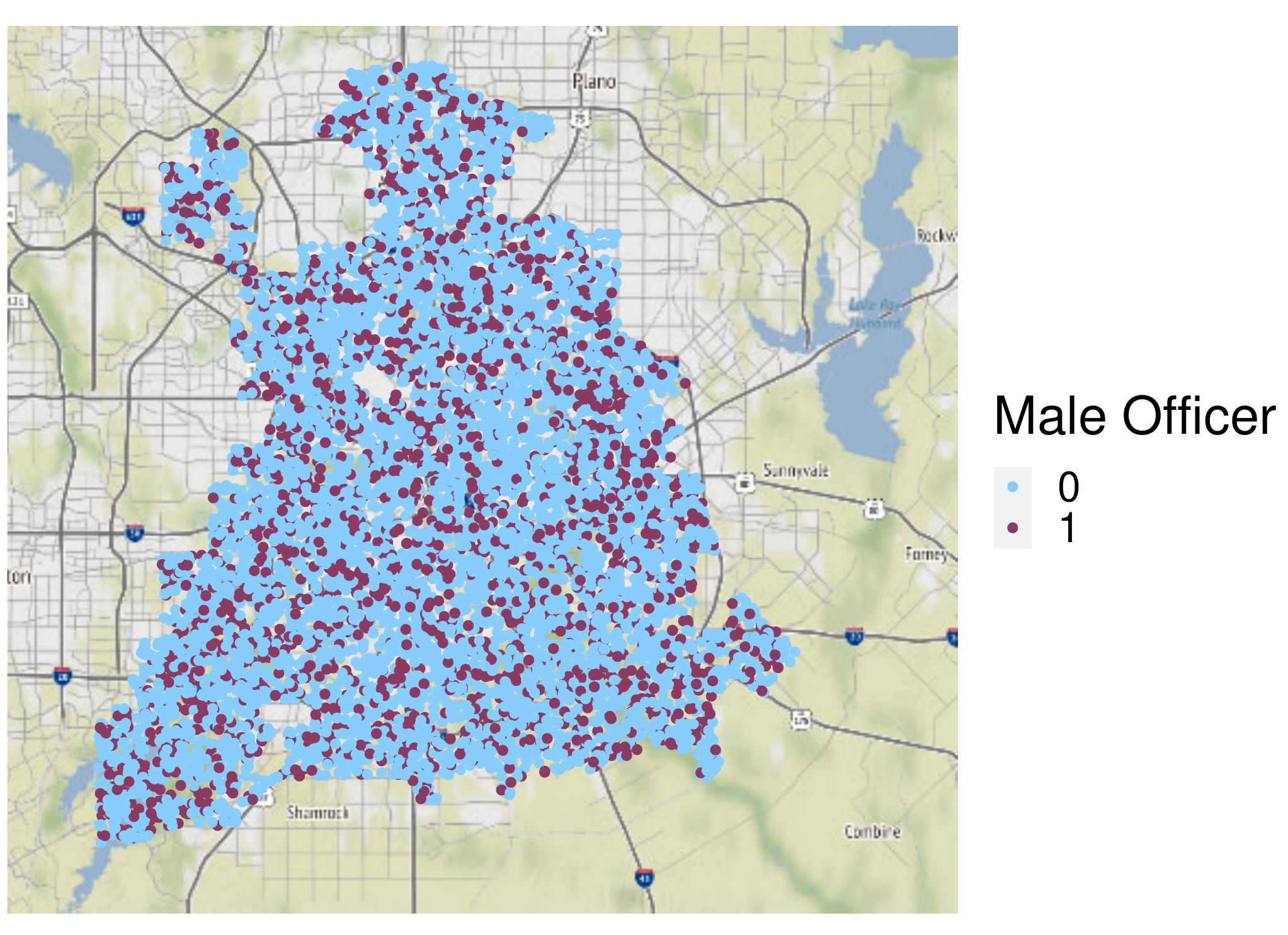} 
}%
\hspace{1.4in}
%\hfill
\subfloat[{Mark-Dependent Thinning:} \\
         $P(\text{keep s}) = \lambda_{\bm{\theta}_k}(s, \nu)/\max(\lambda_{\bm{\theta}_k}(s, \nu))$]{\label{fig:sub-fourth}%
    \hspace{0.3in}
    \includegraphics[height=1.2in]{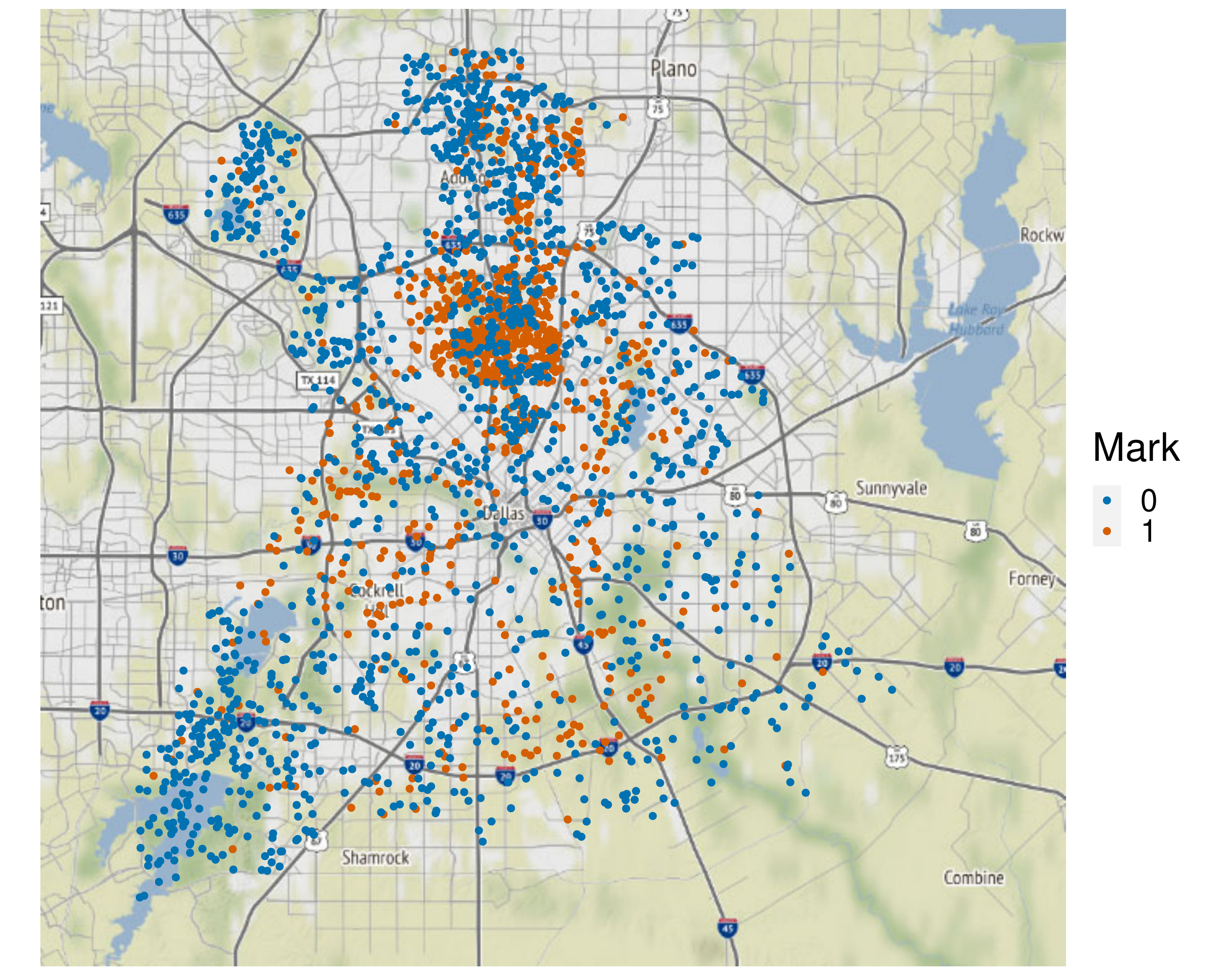}    %
}

\vspace{-10pt}

\caption{Simulation Process for Bivariate Mark Model\label{fig:bivgp_sim}}
\end{figure}

The full simulation algorithm for the bivariate mark model is included in Algorithm \ref{bivar_algorithm} in \ref{sim_alg_appendix}. The steps are depicted in Figure \ref{fig:bivgp_sim}. Of particular concern, we note in the first step the need to calculate $\max(\lambda_{\bm{\theta}_k}(s, \nu))$ for each of $k$ mark levels, where $\lambda_{\bm{\theta}_k}(s, \nu)$ consists of both spatial and nonspatial variables, shown in Figure \ref{fig:sub-first}. This is intuitive for spatial variables but the procedure is not so clear when nonspatial variables are incorporated into the intensity function. In order to calculate $\lambda_{\max}$ for each mark level, we require bounded distributions for the nonspatial variables; this is also required for model inference as shown in detail in \ref{app:int_sec}. In summary, in order to calculate the integral of the intensity over a nonspatial variable that is not categorical ($\int_{V}\exp(\bm{\nu} \bm{\alpha}) dv$) we need to specify an upper and lower bound for the integral. This necessitates bounded distributions. Some choices for bounded distributions include uniform and Beta distributions, and other standard truncated distributions. After calculating $\max(\lambda_{\bm{\theta}_k}(s, \nu))$ for each mark level $k$, we again simulate a homogeneous Poisson Process, but this time with a separate point process for each mark level. Next, we simulate nonspatial variables for each point. Finally, we thin the point process by calculating the intensity at each point and thinning points based on their probability in relation to the maximum intensity to obtain the final point process based on both spatial and nonspatial variables, shown in Figure \ref{fig:sub-fourth}.

\subsection{Predictive Processes}
\label{pred_proc_sec}
\begin{figure}
    \centering
    \includegraphics[width = 0.5\textwidth]{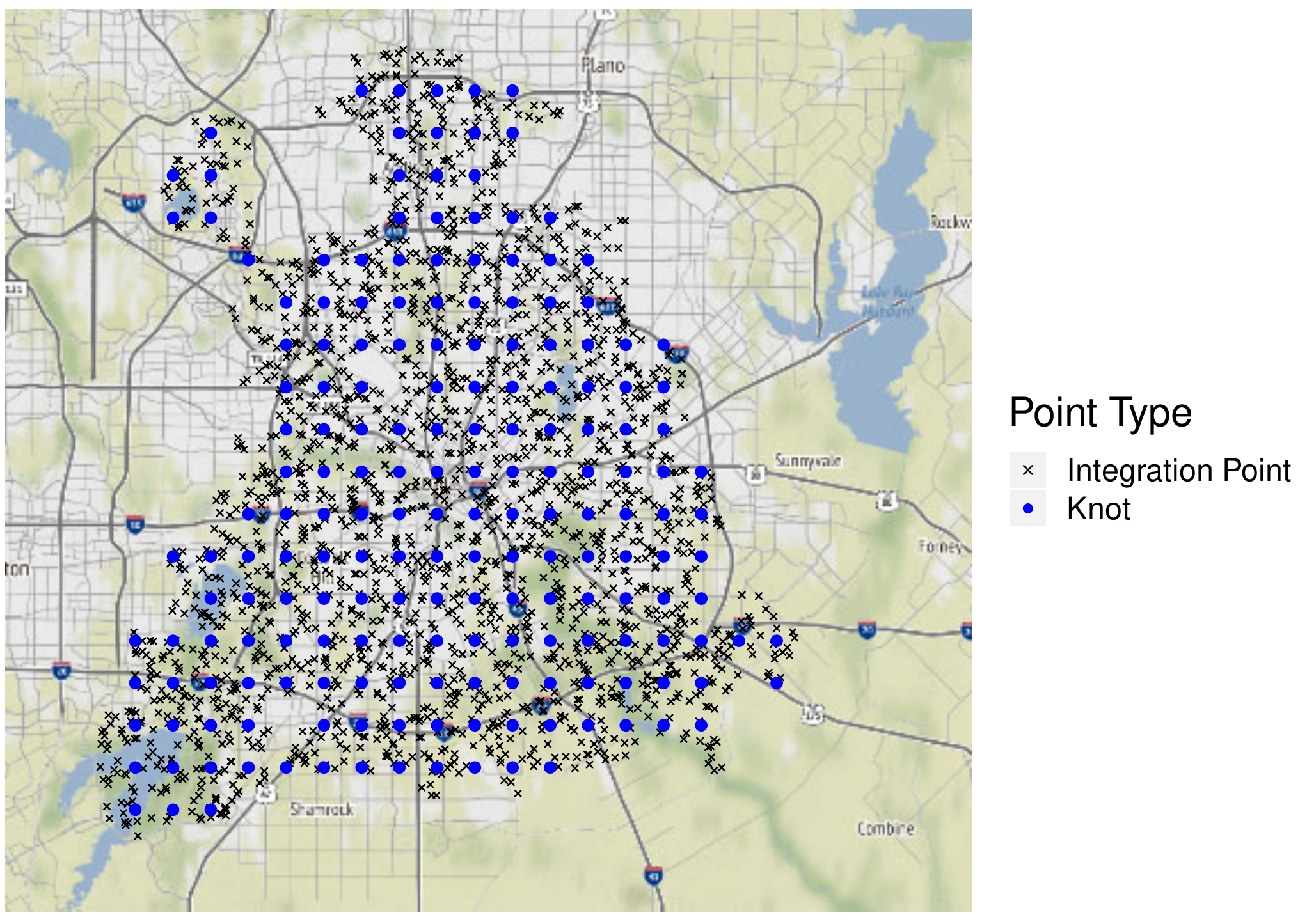}
    \caption{Integration Points and Predictive Process Knots}
    \label{fig:point_type}
\end{figure}

The Gaussian process, $\omega(s)$ needs to be estimated as a finite dimensional random-variable. However, estimating a Gaussian process with a $n\times n$ covariance matrix is computationally intensive. Therefore, we use the predictive process approach advocated for by \cite{banerjee2008gaussian}. We use a set of knots evenly distributed over the window of interest, as in \cite{liang2008analysis}. These knots are shown in Figure \ref{fig:point_type} as the blue circles. If $s^*_1, ...,s^*_n$ are the predictive process knots, and $\bm{\omega^*}$ is a realization of the Gaussian process over the knots, then we transform this realization over the knots to be realizations over both the $s_1,...,s_n$ observed events as well as the integration points, discussed above. The multivariate predictive process realization, $\bm{\Tilde{w}}$, is calculated as follows:

$$\Tilde{w}(s) =  \text{cov}(\omega(s), \bm{\omega^*})\text{var}^{-1}(\bm{\omega^*})\bm{\omega^*}$$ 

For the two-stage model, if $m$ is the number of knots in the predictive process and $n$ is the number of points in the point process, then $\mathit{\textit{cov}(\omega(s), \bm{\omega^*})}$ is the $n \times m$ covariance matrix between the two Gaussian processes. Next, $\mathit{\textit{var}^{-1}(\bm{\omega^*})}$ is the $m \times m$ covariance matrix of the Gaussian process over the knots, $\bm{\omega^*}$ \citep{banerjee2008gaussian, banerjee2014hierarchical}. Then the resulting $\bm{\Tilde{w}}$ is the $n \times 1$ mean-zero predictive process.

For the bivariate mark model, if $k$ is the number of marks and again $m$ is the number of knots in the predictive process and $n$ is the number of points in the point process, then $\mathit{\textit{cov}(\omega(s), \omega^*)}$ is the $nk \times mk$ covariance matrix between the two Gaussian processes. The covariance matrix for the bivariate mark model involves both the exponential correlation function as well as the cross-covariance function. As in the two-stage model, $\mathit{\textit{var}^{-1}(\omega^*)}$ is the $mk \times mk$ covariance matrix of the Gaussian process over the knots, $\bm{\omega^*}$. Then the resulting $\bm{\Tilde{w}}$ is the $nk \times 1$ mean-zero predictive process.

%%%%%%%%%%%%%%%%%%%%%%%%%%%%%%%%%%%%%%%%%%%%
%%%%%%%%%%%%%%%%%%%%%%%%%%%%%%%%%%%%%%%%%%%%
%%%%%%%%%%%%%%%%%%%%%%%%%%%%%%%%%%%%%%%%%%%%
\section{Applications to Simulated Data}
\label{app_sec}

We studied our modeling and inferential approach via simulated examples. Our simulated data closely mirrors the police use of force data from Dallas, Texas, where we create a marked point process with both spatial, using Dallas census tracts, and nonspatial covariates of multiple types. Given several open questions about the use of force dataset, in Dallas, Texas and otherwise, in this manuscript we restrict our attention to the simulated point process data, resembling the use of force data, and the forest fire application. The areal data is not simulated, and is collected from the 2013-2017 American Communities Survey on the census tract level. For spatial variables, we use one variable that is continuous, such as median age, which we denote $z_1(s)$ with a corresponding coefficient $\beta_1$. We also use another spatial variable that is bounded between 0 and 1, such as the unemployment rate or the Herfindahl index, which we denote $z_2(s)$ with a corresponding coefficient of $\beta_2$. For nonspatial variables, we consider one variable that is continuous, to represent a variable such as an officer's tenure on the police force, which is denoted by $\nu_1$ with a corresponding coefficient of $\alpha_1$. We also use a nonspatial variable that is binary, to represent an indicator variable such as male officer, which is denoted $\nu_2$ with a corresponding coefficient of $\alpha_2$. For both the two-stage and bivariate mark models, we consider both log Gaussian Cox process and nonhomogeneous Poisson process models, or models with and without a Gaussian process respectively, in our simulation framework. 
For both the two-stage and bivariate mark models we analyze simulated data generated through thinning procedures, as outlined above, to generate point processes with the appropriate intensity function.

%%%%%%%%%%%%%%%%%%%%%%%%%%%%%%%%%%%%%%%%%%%%
%%%%%%%%%%%%%%%%%%%%%%%%%%%%%%%%%%%%%%%%%%%%
%%%%%%%%%%%%%%%%%%%%%%%%%%%%%%%%%%%%%%%%%%%%
\subsection{Two-Stage Model}
\label{sec:two_stg_fit}

In our first simulated example, we consider a simple model that includes spatial variables in the location determination stage of the model and only nonspatial variables in the mark determination stage of the model, and we do not include a Gaussian process in either stage of the model. Therefore, the first stage of the model is a nonhomogeneous Poisson process (NHPP), rather than an LGCP. We find that we are able to recover the logistic regression estimates for the nonspatial variables.  
The coefficients for the spatial variables are also recovered well when we do not include a Gaussian process. 

\begin{figure}[ht]
    \centering
    \includegraphics[width = 0.95\textwidth]{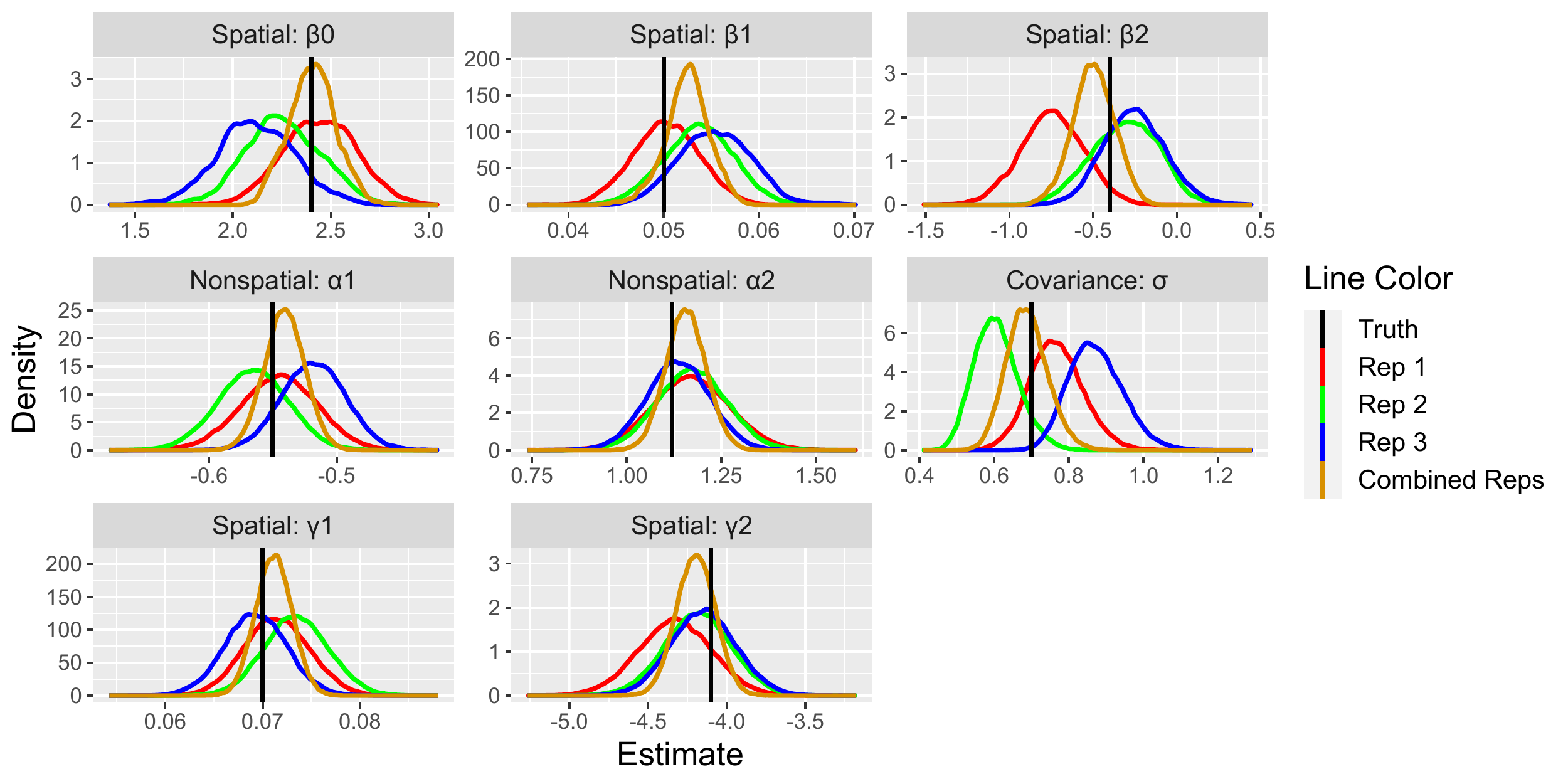}
    \caption{We simulate data from the two-stage model, with a Gaussian process in the first stage of the model. The posterior densities of all the parameters for three replicates of this data are shown in red, green, and blue. When the replicates are fit simultaneously, by multiplying the likelihoods, the posterior densities for the parameters are shown in orange.}
    \label{fig:two_stg_wgp_repsatonce}
\end{figure}

When we add a Gaussian process to the first stage of the model to form an LGCP, we find that we are able to recover regression parameters associated with spatial and nonspatial variables again reasonably well. However, the estimate of $\sigma$ from the covariance of the Gaussian process is not as accurate. It is often difficult in LGCP-type models to tease apart information about these parameters from a single realization of the point process. To study this further, and because there are situations where replications are available, we consider the situation where there are multiple realizations, or replicates, of the point process, using the same parameters. 
In Figure \ref{fig:two_stg_wgp_repsatonce}, the red, blue, and green lines represent the inferred values using just a single realization, that is, when we consider each of the three replicates separately. We see that they range rather widely, especially for the intercept, $\beta_0$ and the covariance parameter, $\sigma$. When we repeat this analysis with varied parameter settings, we find similar results where parameters are estimated relatively well and replicates improve our ability to accurately estimate parameters.

However, when we infer parameters based on all three replicates at once, 
 we find that we better estimate all the parameters, including $\sigma$ and $\beta_0$ (orange curves in Figure \ref{fig:two_stg_wgp_repsatonce}). 
  The likelihood function here is simply the product of the three likelihoods in Equation \ref{eq:two_stage_mod} across the individual replicates, assuming independence between replicates.

%%%%%%%%%%%%%%%%%%%%%%%%%%%%%%%%%%%%%%%%%%%%
%%%%%%%%%%%%%%%%%%%%%%%%%%%%%%%%%%%%%%%%%%%%
%%%%%%%%%%%%%%%%%%%%%%%%%%%%%%%%%%%%%%%%%%%%

\subsubsection*{Two-stage model with cross-correlation}
We also simulate from and fit the more general version of the two-stage model shown in Equation \ref{eq:dep_model} to study the impact of the flexible assumption of dependence across the two stages of the model. 
We test all possible combinations of simulated data and models fitted to that data from Table \ref{tab:model_type}, including fitting simple models to more complex data, and complex models to simple data. This allows us to test whether our modeling procedure shows an improved fit and can recover parameters from the more complicated model where there is dependence between the two stages.

\begin{table}
\centering
\begin{tabular}{|l|rrrr|}
 \hline
 
 & \multicolumn{4}{c|}{Simulated Model (Truth)} \\
 Fitted Model & Model 1 & Model 2 & Model 3 & Model 4 \\
  \hline
 Model 1 & -17,954.09 & -17,583.53 & -8,917.32 & -22,075.13 \\
Model 2 & -17,955.55 & -18,069.73 & -9,817.12 & -23,807.32  \\
 Model 3 & {\bf-17,989.75}  & {\bf-18,231.51} & -9,860.67 & -23,930.88 \\
 Model 4 & -17,975.52 & -18,213.73 & {\bf-9,892.37} & {\bf-23,968.50}  \\
   \hline
\end{tabular}
\caption{\label{tab:tab03b} Model Comparison Two-Stage Model, Simulated Data vs Model Fit (WAIC). Rows represent different models that were fit to the data and columns represent the model that the data was simulated from. Bold values indicate best model fit.} %\\
\end{table}

We present the results in Table \ref{tab:tab03b}. Model comparison is conducted with the Watanabe–Akaike information criterion (WAIC) which is a Bayesian approach to balancing model fit and model complexity \citep{watanabe2010asymptotic}. We find that the bivariate and univariate (univariate GP in both stages) models have the lowest WAIC and therefore best model fit, regardless of the form of the simulated data. The WAIC values tend to be similar between the two models. When the second stage of the model does not include a GP (Models 1 and 2), the model with a univariate GP in both stages provides the best fit. When the second stage of the model includes a GP (Models 3 and 4), regardless of dependence in the GP, the model with a bivariate GP across stages provides the best fit. In Table \ref{tab:tab03b}, we include the simulated case of the bivariate GP model where $\rho$ is 0.9.

We also test the comparison of models for different values of $\rho$. We focus on the case where data is simulated from the two-stage model with dependent Gaussian processes across stages (model 4). We find that WAIC values indicate a similar model fit for both the model with a bivariate GP across stages and the model with a univariate GP in both stages of the model, regardless of the value of $\rho$ that is used.
However, we find that these models with spatial dependence (models 3 and 4) are distinguished effectively from models without spatial dependence in one or both stages of the model (models 1 and 2).
From this analysis, we suggest that it may be difficult at times to distinguish between models 3 and 4 using WAIC, where in model 4 we add dependence between the two stages of the model, using the $\rho$ parameter. Even if the independent model is indicated as the better fit through WAIC, there may be dependence across stages.

\begin{figure}[ht]
    \centering
    \includegraphics[width = 0.75\textwidth]{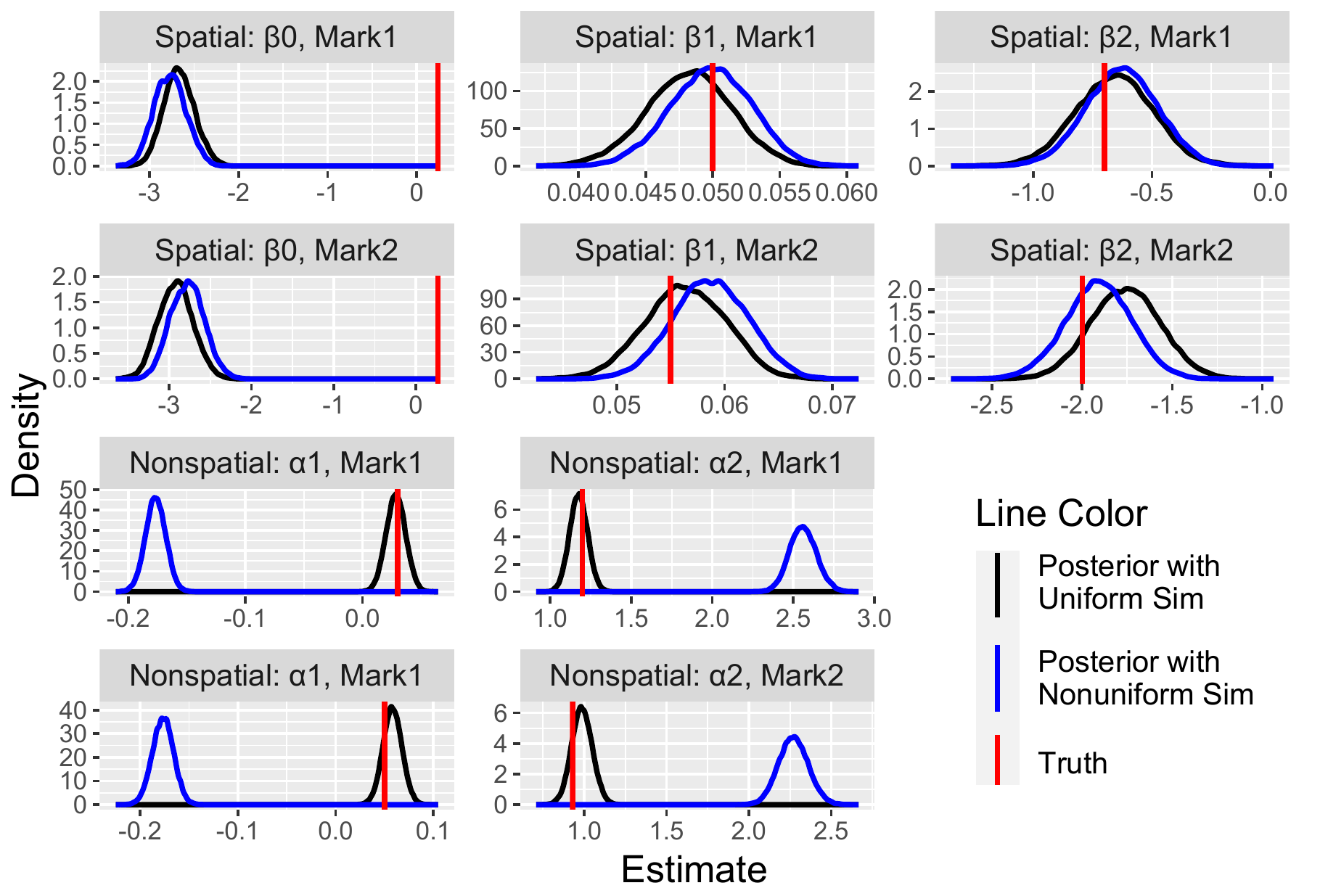}
    \caption{We simulate data from the bivariate mark model, without a Gaussian process, and compare the results for the when the nonspatial variables are simulated from uniform or nonuniform distributions. The posterior densities for all parameters are shown in black when we include nonspatial variables simulated from uniform distributions and blue for nonspatial variables simulated from nonuniform distributions. The true parameters from the simulation are shown in red.}
    \label{fig:nonsp_nogp}
\end{figure}

\subsection{Bivariate Mark Model}
\label{sec:bivmodel_sim}

We test many different cases of the bivariate mark model with both spatial and nonspatial variables. From our results, we suspect that there are identifiability issues between the nonspatial variables and the intercept, as the intercept is not recovered well when we include nonspatial variables, even when we do not include a Gaussian process. In Step 5 of the bivariate mark model's simulation algorithm as described in Algorithm \ref{bivar_algorithm} in the Appendix, we simulate nonspatial variables from pre-specified distributions. We find that we are only able to recover nonspatial variables when they are simulated from uniform/equal probability distributions. As shown in the Appendix in Equation \ref{nonsp_integral}, bounds on continuous distributions are required in order to calculate the integral of the intensity function. However, we find that even when we simulate from a bounded distribution that is not uniform, we cannot successfully recover the nonspatial variable coefficients. When we draw the realizations of the nonspatial variables from a uniform distribution, we are able to effectively recover the nonspatial variable coefficients, but still are unable to recover the intercept. When the nonspatial variables are binary, we also find that the only way to recover nonspatial coefficients is to simulate with equal probability from 0 and 1.

We illustrate our findings in Figure \ref{fig:nonsp_nogp}, where the models are estimated without a Gaussian process through nonhomogeneous Poisson processes rather than LGCPs. The black line shows the posterior distributions of the nonspatial parameters where the associated variables are simulated from a uniform distribution for the continuous variable ($\nu_1$) and a Bernoulli distribution with equal probability on 0 and 1 for the binary variable ($\nu_2$). We see that, with the exception of the intercept, the spatial and nonspatial variables are recovered reasonably well in most cases. We note that even though the variables are simulated from these uniform/equal probability distributions, our simulation process relies on a thinning procedure, as described in Algorithm \ref{bivar_algorithm} in the Appendix, and the distributions of the nonspatial variables after thinning no longer resemble uniform/equal probability distributions. The blue line in Figure \ref{fig:nonsp_nogp} shows the posterior distributions when the nonspatial variables are simulated from non-uniform distributions and Bernoulli distributions with unequal probability. The nonspatial variables are not estimated well, and the spatial variables are recovered similarly well to the case above. We have tested many choices for the prior distribution of the regression coefficients, especially the intercept terms, and these results remained consistent.

\begin{figure}
    \centering
    \includegraphics[width = 0.75\textwidth]{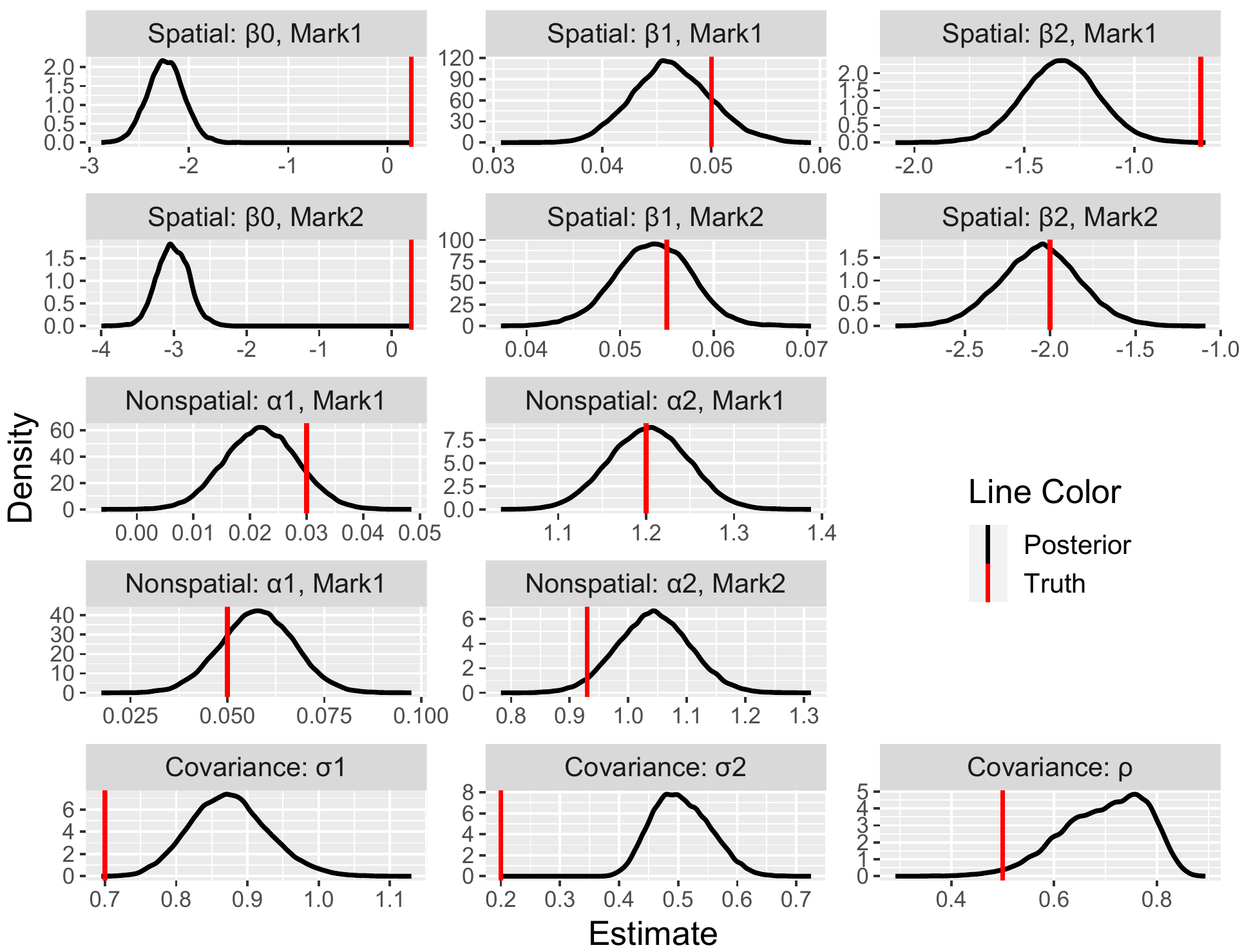}
    \caption{We simulate from a bivariate mark model with a bivariate Gaussian process between the two simulated levels of force. The posterior densities for all parameters are illustrated in black, and the true parameter from the simulation is shown in red.}
    \label{fig:bivgp_wgp}
\end{figure}

We illustrate our LGCP results, with a Gaussian process added to the intensity function, in Figure \ref{fig:bivgp_wgp}. When we add the bivariate Gaussian process to this model, with cross-covariance $\Lambda$, we find once again that we are unable to recover the intercept parameters, $\beta_0$ for either mark. We are able to recover both spatial ($\bm{\beta}$) and nonspatial ($\bm{\alpha}$) variable coefficients relatively well, when the $\bm{\nu}$ are once again simulated from uniform/equal probability distributions. The main inferential challenge in this method is estimating the parameters of the bivariate Gaussian process. We find that we are able to recover a difference between $\sigma_1$ and $\sigma_2$, namely that $\sigma_2$ is smaller than $\sigma_1$, but we do not recover the true value well. We do not recover the parameter $\rho$ well, as our point estimate for $\rho$ is around 0.7, while the true value is 0.5.

%\begin{comment}
%\end{comment}

In summary, these simulated examples have shown that our two-stage model provides accurate and intuitive estimation of parameters in both the location and the mark determination stages of the model. Replicates can be valuable, if available for an application, especially in recovering the parameter(s) of the covariance function. When incorporating dependence across the two stages of our model, it is often difficult to distinguish between the case where the Gaussian processes are independent or dependent using model choice metrics such as WAIC. Therefore, due to the computational costs of estimating the case of dependence across stages, users may choose to estimate the case where there are independent Gaussian processes in both stages. For the bivariate mark model, we have shown through the simulated examples that it is difficult to recover intercept parameters and there are constraints placed on the distributions of nonspatial variables, given the placement of nonspatial variables in the spatial intensity function.

%%%%%%%%%%%%%%%%%%%%%%%%%%%%%%%%%%%%%%%%%%%%
%%%%%%%%%%%%%%%%%%%%%%%%%%%%%%%%%%%%%%%%%%%%
%%%%%%%%%%%%%%%%%%%%%%%%%%%%%%%%%%%%%%%%%%%%
\section{Application to Castilla-La Mancha Fire Data}
\label{sec:fire_data}
Next, we apply the two-stage model to data on forest fires from the Castilla-La Mancha region of Spain, shown in Figure \ref{fig:firedata}. The fire data used to evaluate the model is the \textit{clmfires} forest fire data from the spatstat package in R \citep{spatstat}. Nonspatial variables with the fire data indicate the cause of the fire and whether the fire occurred in the summer season. Spatial variables include elevation and other characteristics of the land, such as land cover and slope, which are available on a fine grid. We are also interested in characterizing which factors impact the amount of land burned in each forest fire. We consider 3,657 fires from 2004 to 2007. Data collected before 2004 were recorded with less precision than after 2004, so we exclude these cases. We also exclude fires that had ``other'' listed as the cause of the fire, as we are interested in comparing intentional fires to those caused by lightning or an accident. In the first stage of the model, we test the following spatial variables and their impact on determining the locations of forest fires in the Castilla-La Mancha region of Spain: indicator variable for if the location is in a forest, elevation, and the slope at that location (note that the spatial covariate data may need to be transposed after loading it from the spatstat package). In the second stage of the model, we use the following nonspatial variables: an indicator variable for if the fire was set intentionally (compared to being set by lightning or an accident) and an indicator variable for if the fire occurred in the summer season. We combine the spatial and nonspatial variables listed above to determine their potential impact on the mark of the point process, or the amount of land burned, in hectares, by a given fire. Therefore, this model is slightly modified from the two-stage model shown in Equation \ref{eq:dep_model}, where the second stage is now linear regression, where previously it was logistic regression in the case of simulated police use of force incidents. In the second stage, we use the log burned area as the mark and therefore response in the linear regression. This is an advantage of our two-stage model, as it allows the user to consider many different kinds of marks, including continuous and categorical, whereas the bivariate mark model proposed by \cite{liang2008analysis} allows only for categorical marks and therefore the bivariate mark model cannot be applied to the \textit{clmfires} data.

\cite{alba2018homogeneity} analyze the \textit{clmfires} data by comparing two different fire seasons per year. The authors find it is important to consider whether fires occur in the summer season during some years, motivating the use of this nonspatial covariate in our model. \cite{myllymaki2020testing} use the \textit{clmfires} data to model local covariate effects. Specifically, they incorporate elevation and an indicator variable for whether the fire occurred in a forest in the spatial model, motivating our use of these variables as well. \cite{dvovrak2020nonparametric} develop nonparametric methods to study the relationship between points, marks, and covariates. Through the use of test statistics, they conclude that there is dependence between the points, marks, and covariates that they consider (the mark is the area burned and the covariate is elevation). The authors consider only fires in 2007, and when studying the relationship between points and (spatial) covariates alone, they find no relationship between elevation and locations of fires, which will be tested in the first stage of our model.

We begin our analysis of the 3,657 from 2004 to 2007 in the \textit{clmfires} dataset by conducting preliminary tests of correlation between different combinations of the points, marks, and covariates. We begin with tests based on random shifts, as introduced by \cite{dvovrak2020nonparametric} to test the dependence between spatial covariates and the points (denoted as a Point-Covariate test, or P-C test). As in \cite{dvovrak2020nonparametric}, we perform 999 shifts with radius up to 150km for this dataset. The region is approximately 450km $\times$ 450km. After the variance correction which accounts for points being moved outside of the observation window, the p-values for the elevation, slope, and indicator variable for forest are 0.73, 0.052, and 0.59 respectively. When these tests are bound together by the Bonferroni test, as suggested in \cite{dvovrak2020nonparametric}, we find that none of the spatial covariates impact the distribution of points. This will be further investigated in the first stage of our model, where we have developed a parametric model to study the relationship between all three covariates and the distribution of points.

Next, we investigate the geostatistical marking model, which implies that the point process and marks are independent. We conduct visual tests using the functions $E(r)$ and $V(r)$, defined by \cite{schlather2004detecting} as the conditional mean and variance of the mark attached to a typical random point, given that there is a further point of the process a distance $r$ away. Given geostatistical marking, the functions $E(r)$ and $V(r)$ would be constant at any distance $r$. For the fire data, we find that $E(r)$ and $V(r)$ tend to increase with distance $r$, providing evidence against geostatistical marking. The mark-weighted K function \citep[cf.][]{guan2006tests} and corresponding simulation envelopes for random labelling (implemented through the $Kmark$ function in R) also show evidence against random labelling, showing dependence between the points and marks. This is consistent with the findings on the subsetted fire data from \cite{dvovrak2020nonparametric}. This rejection of geostatistical marking suggests the need to investigate the spatial dependence between marks and points further. The Gaussian process in the second stage of our two-stage model is a useful tool to do so.

Finally, we conduct a combined test of dependence between the marks and spatial covariates through the random shift test developed by \cite{dvovrak2020nonparametric}. We use the Kendall's correlation coefficient test with variance correction and find the following p-values for elevation, slope and indicator variable for forest: 0.04, 0.014, and 0.002. When the tests are bound together with the Bonferroni correction, we find that with a significance level $\alpha = 0.05/3 = 0.017$, only the forest indicator variable is significant when analyzing the relationship between marks and the covariate.

These tests are powerful exploratory tools to analyze the dependence in the data between points, marks, and the covariates. Our two-stage parametric model allows us to investigate these relationships in more detail. The two-stage model allows us to analyze the relationship between points and covariates in the first stage, covariates and marks in the second stage, and the relationship between all three using the dependent Gaussian processes across stages. It allows us to incorporate unobserved spatial patterns into both the location and mark determination processes through Gaussian processes. We investigate our detailed findings using the two-stage model next.

We fit four models in the two-stage framework to the fire data. The first model considers a non-homogeneous Poisson process (NHPP) in the first (location-determination) stage and standard normal error in the second (mark-determination) stage. In the second model, the first stage has a Gaussian process and the second stage has standard normal error. In the third model, there are univariate Gaussian processes in both stages of the model, with no dependence between them. In the final model, we consider dependent Gaussian processes in the first and second stages.
 The results for model comparison are presented in Table \ref{tab:tab04b}. We find that the lowest WAIC is for the model which includes a univariate GP in both stages of the model, with no dependence between the stages. This is followed by the bivariate Gaussian process model, with dependence between the stages.

\begin{table}
\centering
\begin{tabular}{|lllr|}
 \hline
 Model Type & First Stage & Second Stage &  WAIC  \\ 
  \hline
    Model 1 & NHPP & $\epsilon \sim N(0, \sigma^2)$ & -4,317.87 \\
    Model 2 & Univ. Gaussian Process & $\epsilon \sim N(0, \sigma^2)$ & -7,070.73  \\
    Model 3 & Univ. Gaussian Process & Univ. Gaussian Process & {\bf-7,497.39} \\
    Model 4 & \multicolumn{2}{c}{ Biv. Gaussian Process} & -7,353.62 \\ 
   \hline
\end{tabular}
\caption{\label{tab:tab04b} Model Comparison Two-Stage Model, Castilla-La Mancha Spain Forest Fire Data} %\\
\end{table}

\begin{comment}
\begin{table}
\centering
\begin{tabular}{|lllr|}
 \hline
 Model Type & First Stage & Second Stage &  WAIC  \\ 
  \hline
    Model 1 & NHPP & $\epsilon \sim N(0, \sigma^2)$ & -3,866.11 \\
    Model 2 & Univ. Gaussian Process & $\epsilon \sim N(0, \sigma^2)$ & -2,328.69  \\
    Model 3 & Univ. Gaussian Process & Univ. Gaussian Process & {\bf-7,472.51} \\
    Model 4 & \multicolumn{2}{c}{ Biv. Gaussian Process} & -5,657.39 \\ 
   \hline
\end{tabular}
\caption{\label{tab:tab04b} Model Comparison Two-Stage Model, Castilla-La Mancha Spain Forest Fire Data} %\\
\end{table}
\end{comment}

Next, we analyze the effects of the variables included in both stages of the model. We show the results for Model 3, or the model with a univariate GP in both stages, in Table \ref{tab:tab05_sub}, and the results for all models in the Appendix, Table \ref{tab:tab05_full}. Values in bold in the table are coefficients whose credible intervals do not include 0. For Model 3, which was indicated as the best fit through WAIC, we find that the indicator variable for forest cover and elevation both have negative impacts on the spatial intensity of fires, while the slope has a positive effect on the spatial intensity. This would indicate that at locations with higher elevation, forested regions, and regions with a smaller slope, we are less likely to see fires. These trends are consistent for the first stage of the model in all model types, shown in Table \ref{tab:tab05_full}, except for the NHPP where the forest indicator variable is found to have no effect.

In the second stage of the model, we find for Model 3  that the forest indicator variable and slope have negative effects on the amount of area burned, and slope has no effect. For nonspatial variables, we find that the binary variable indicating that the fire was intentional has no effect, and the summer indicator variable has a negative effect. Therefore, we expect smaller fires during the summer and in regions that have forest cover and have higher elevation. The negative impact of elevation on area burned is consistent with exploratory analysis conducted by \cite{dvovrak2020nonparametric}.
We find that the results are dependent on whether a Gaussian process is included in the second stage of the model. When we do not include a GP in the second stage (Models 1 and 2), we find that both slope and the indicator variable for whether the fire was intentionally set have impacts on the amount of area burned where these have no effect with a Gaussian Process (Models 3 and 4).  

Finally, we consider the dependence parameter, $\rho$, in the fourth model which includes a bivariate Gaussian process. The credible interval for $\rho$ does not include 0, which indicates dependence between the two stages of the model. The $\rho$ parameter is estimated to be negative, indicating negative spatial structure in variables unaccounted for between the location and mark determination processes. Through the WAIC, we have indicated that the model without WAIC is a better fit and our simulation studies show that WAIC may indicate the independent GP model (Model 3) even when there is dependence between the stages. Therefore, further investigation into the spatial structures in the processes that determine where forest fires happen and the severity of these fires is necessary.

\begin{comment}
\begin{table}
\centering
\begin{tabular}{|llrr|}
\hline
& & Stage 1 & Stage 2\\
\hline
&Variable & Mean (95\%CI) & Mean (95\%CI)  \\ 
 \hline
 \multirow{5}{*}{\rotatebox[origin=c]{90}{Spat Var}} &$\beta_0$ (intercept) & {\bf4.1115} (3.9397, 4.2862)  & 0.0571 (-0.3231, 0.4236) \\ 
 &Forest (ind.) & -0.0494 (-0.1531, 0.0537)  & 0.0479 (-0.1735, 0.2697)\\ 
 &Elevation & {\bf-7$\times10^{-04}$} (-9$\times10^{-04}$, -5$\times10^{-04}$) & {\bf-8$\times10^{-04}$} (-0.0013, -4$\times10^{-04}$)\\ 
 &Slope & {\bf0.0089} (1$\times10^{-04}$, 0.0171) & -0.0121 (-0.0276, 0.0028) \\ 
&$\sigma_1$ or $\sigma_2$ &5.3444 (4.8574, 5.8351) &  6.694 (5.6753, 7.7473) \\ 
& $\sigma$ (iid) & & 2.1898 (2.1389, 2.2415) \\
 \hline
 \multirow{3}{*}{\rotatebox[origin=c]{90}{Nonsp}}& &&\\ 
 & Intentional (ind.) & & 0.0318 (-0.1314, 0.2)  \\
 &Summer (ind.) & & {\bf-0.2318} (-0.3787, -0.0807)   \\  

\hline
\end{tabular}
\caption{\label{tab:tab05_sub} We present the posterior mean as well as the 95\% Credible Intervals for all covariates in our analysis of the forest fire data in the Castilla-La Mancha region of Spain. Values in bold indicate coefficients whose credible intervals do not contain 0. We present the estimates for Model 3 (univariate GP in both stages) above and the full results in the Appendix (Table \ref{tab:tab05_full}).} %\\
\end{table}
\end{comment}

\begin{table}
\centering
\begin{tabular}{|llrr|}
\hline
& & Stage 1 & Stage 2\\
\hline
&Variable & Mean (95\%CI) & Mean (95\%CI)  \\ 
 \hline
 \multirow{5}{*}{\rotatebox[origin=c]{90}{Spat Var}} & Intercept & {\bf4.492} (4.3008, 4.6831)  & {\bf1.4333} (1.0825, 1.7868)  \\ 
 &Forest (ind.) & {\bf-0.1929} (-0.3005, -0.0825)  & {\bf-0.4988} (-0.7337, -0.2745)\\ 
 &Elevation & {\bf-0.0013} (-0.0015, -0.0011) & {\bf-0.0027} (-0.0031, -0.0022)\\ 
 &Slope & {\bf0.026} (0.0186, 0.033) & 0.0074 (-0.0088, 0.0228) \\ 
&$\sigma_1$ or $\sigma_2$ & {\bf5.1529} (4.6399, 5.6641) &  {\bf3.4521} (2.7675, 4.2067) \\ 
& $\sigma$ (iid) & & {\bf2.1906} (2.1399, 2.2416) \\
 \hline
 \multirow{3}{*}{\rotatebox[origin=c]{90}{Nonsp}}& &&\\ 
 & Intentional (ind.) & & 0.0286 (-0.1358, 0.191)   \\
 &Summer (ind.) & & {\bf-0.2191} (-0.3678, -0.0747)    \\ 
\hline
\end{tabular}
\caption{\label{tab:tab05_sub} We present the posterior mean as well as the 95\% Credible Intervals for all covariates in our analysis of the forest fire data in the Castilla-La Mancha region of Spain. Values in bold indicate coefficients whose credible intervals do not contain 0. We present the estimates for Model 3 (univariate GP in both stages) above and the full results in the Appendix (Table \ref{tab:tab05_full}).} %\\
\end{table}
%\end{comment}

%%%%%%%%%%%%%%%%%%%%%%%%%%%%%%%%%%%%%%%%%%%%
\section{Discussion}
\label{sec_disc}
Through this study, we have demonstrated the utility of the two-stage model in estimating the relationship between spatial and nonspatial variables and marked point process outcomes. Through the two applications discussed in this paper, we have also shown how the model can be applied to multiple types of marks, such as continuous and categorical. Specifically, with the police use of force application, we are interested in nonspatial variables such as officer and citizen race/gender, officer tenure, citizen resistance, and the number of citizens and officers involved. Spatial variables include community factors, such as median income and neighborhood diversity. These spatial and nonspatial variables could be related to the mark, or the level of force that is used by the police officer. Through a simulated example created to closely resemble police use of force data, we have shown that this two-stage model provides interpretable and accurate estimates of relationships between covariates and both the intensity and mark of point processes. In the simulated example, we have shown that replicates can be helpful in recovering parameters, especially the covariance parameters of the Gaussian process. There may be opportunities in some applications to treat segments of the data as replicates, for instance annual data could be treated as replicates. In many cases, however, point processes can be expected to vary over time and space, and therefore it is difficult to consider any of the data as replicates. 

For the forest fire data, we are interested in determining which factors play a role in both determining the location of forest fires and the amount of burned area at the fire locations. We show that this model provides a clear interpretation of nonspatial and spatial variables and the relationship with these marked point process outcomes by splitting the model into two stages: location determination and mark determination. We show that the factors impacting the location of fires and the factors impacting the amount of burned area at those locations are not necessarily the same, and accounting for spatial dependence when determining the impact of variables on the mark is important.  

In addition to being flexible to many types of marks, our two-stage model could also be expanded to include multiple marks, with an additional stage added to the model with each additional mark of interest. Adding Gaussian processes into the location or additional mark determination stages of the model, as motivated by the application, would add computational complexity but could allow for exploration of the dependence between marks or between individual marks and the location determination process. A variety of marking models can be incorporated into the second stage, including more explicit models for the dependence between the spatial intensity and marking process, such as that of \cite{myllymaki2009conditionally}.

We anticipate that this new two-stage model could be widely applicable to other types of datasets. For example, \cite{liang2008analysis} studies the impact of nonspatial variables on cancer. The authors parameterize their results of the bivariate mark model to show fixed effects on one mark, and differential effects for the other mark, where they hope to show if the variable has a different impact on the two cancers. In the first stage of the model, we would determine if the spatial variables contribute to an increased intensity of the outcome of interest, in \cite{liang2008analysis}'s case this would be cancer. In the second stage of the model, we could incorporate spatial and/or nonspatial variables to determine the perhaps differential impact of the variables on the different marks, or types of cancer. The dependence structure in the bivariate Gaussian process could be used to test for dependence between the location-determination and the marking processes.

In the future, we would like to apply this model to use of force data provided through the Police Data Initiative and Open Data city websites to conduct an in-depth analysis of the impact of both spatial and nonspatial variables on police use of force, and how these trends compare across cities and jurisdictions. 

\section{Acknowledgements}
The authors would like to thank Aleksandra Slavkovi\'c, Ephraim Hanks, and Corina Graif for helpful conversations. This project was supported by Award No. 2020-R2-CX-0033, awarded by the National Institute of Justice, Office of Justice Programs, U.S. Department of Justice. The opinions, findings, and conclusions or recommendations expressed in this publication/program/ exhibition are those of the author(s) and do not necessarily reflect those of the Department of Justice.

%%%%%%%%%%%%%%%%%%%%%%%%%%%%%%%%%%%%%%%%%%%%
%%%%%%%%%%%%%%%%%%%%%%%%%%%%%%%%%%%%%%%%%%%%
%%%%%%%%%%%%%%%%%%%%%%%%%%%%%%%%%%%%%%%%%%%%

\appendix

%%%%%%%%%%%%%%%%%%%%%%%%%%%%%%%%%%%%%%%%%%%%
%%%%%%%%%%%%%%%%%%%%%%%%%%%%%%%%%%%%%%%%%%%%
%%%%%%%%%%%%%%%%%%%%%%%%%%%%%%%%%%%%%%%%%%%%
\section{Computational Details}
\subsection{Integration and Nonspatial Variables} 
\label{app:int_sec}
\vspace{12pt}

The integral presented in Equation \ref{bvgp_likelihood} cannot be evaluated explicitly and must therefore be approximated. If we omit the mark, the integral to be approximated is as follows: 

$$-\int_{W}\int_{V} \lambda_{\bm{\theta}}(s, v)dvds$$

When we consider an intensity function of the form $log(\lambda_{\bm{\theta}}(s)) = \bm{z(s)}'\bm{\beta} +\bm{\nu}'\bm{\alpha} + \omega(s)$, where there is no interaction between the spatial and nonspatial variables, then the integral can be separated by spatial and nonspatial variables, shown below. 

$$-\int_{W}\int_{V} \lambda_{\bm{\theta}}(s, v)dvds = -\int_{W}\exp(\bm{z(s)}'\bm{\beta} + \omega(s))ds\int_{V}\exp(\bm{\nu}' \bm{\alpha}) dv$$

As in \cite{liang2008analysis}, if a nonspatial variable in $\bm{\nu}$ is continuous, we must specify an upper and lower bound for the covariate, say $\nu_u$ and $\nu_l$, respectively. However, if $\bm{\nu}$ is categorical or discrete, it is less clear how to incorporate these nonspatial variables into the integral. \cite{liang2008analysis} and \cite{banerjee2014hierarchical} both indicate that integrals over discrete variables should be replaced by sums but also that when $\bm{\nu}$ is categorical, than the integration is only over the spatial domain, $W$ \cite[p.~234]{banerjee2014hierarchical}. In our application, we are interested in nonspatial categorical (binary) variables such as officer gender, citizen gender, and if the officer and the citizen are the same race or not. We find that when we do not incorporate these categorical variables into the integral of the intensity function, the estimates for $\bm{\alpha}$ are not accurate.

Therefore, we outline how to evaluate this integral while incorporating nonspatial variables. We provide an example where $\bm{\nu_1}$ is continuous over a bounded distribution from $\nu_{1,l}$ to $\nu_{1,u}$ and $\bm{\nu_2}$ is a binary indicator variable. The integral over $\bm{\nu_2}$ is the sum of the function evaluated at 0 and 1, as suggested by \cite{banerjee2014hierarchical} where the authors suggest that discrete variables are integrated with sums. This is easily extendable to additional continuous and binary variables.

\begin{equation}
    \begin{split}
         \int_{V}\exp(\bm{\nu} \bm{\alpha}) dv = & \int_{V_1}\int_{V_2}\exp(\nu_1\alpha_1 + \nu_2\alpha_2)dv_1dv_2 \\
         = &  \frac{\exp(\nu_1\alpha_1)}{\alpha_1}\Big|_{\nu_{1,l}}^{\nu_{1,u}} \times \sum_{\nu_2 \in {0,1}} \exp(\nu_2\alpha_2) \\
         = & \Big(\frac{\exp(\nu_{1,u}\alpha_1)-\exp(\nu_{1,l}\alpha_1)}{\alpha_1}\Big) \times (\exp(\alpha_2) + 1)
    \end{split}
    \label{nonsp_integral}
\end{equation}

%%%%%%%%%%%%%%%%%%%%%%%%%%%%%%%%%%%%%%%%%%%%
%%%%%%%%%%%%%%%%%%%%%%%%%%%%%%%%%%%%%%%%%%%%
%%%%%%%%%%%%%%%%%%%%%%%%%%%%%%%%%%%%%%%%%%%%
\subsection{Integration Methods}
\label{app_sec:integ_methods}
We investigated many different integration methods in this study. It is important to get an accurate measure of the integral of the intensity function in order to accurately conduct inference. First, we find it is important to include at least one point in each areal unit (in our case, census tracts) as the intensity function depends on the evaluation of the function over each areal unit. The spatial variation of the intensity function depends only on the census tracts and the Gaussian process. Therefore, it is important to include more than one point in large census tracts to capture the variation in the Gaussian process. As in \cite{liang2008analysis}, we set the number of integration points per census tracts as a minimum of one and otherwise proportional to the area. We place these points randomly in the given census tract after determining the number of points to be placed in that census tract. These integration points are shown in Figure \ref{fig:point_type} as the black crosses. We test different integration methods over the region, such as Delaunay triangulation, and we find that a quasi-Monte-Carlo estimate over the integration points produces similar estimates, and we proceed with a quasi-Monte-Carlo estimate for the integral.

\section{Bivariate Mark Model Simulation Algorithm}
\label{sim_alg_appendix}

Below, we include the detailed simulation algorithm for the bivariate marked point process (Algorithm \ref{bivar_algorithm} in \ref{sim_alg_appendix}). The simulation algorithm for the two-stage model is shown in Section \ref{sec:two_stg_desc}. Both algorithms are structured based on the thinning technique, where we start with a homogeneous Poisson Process and we thin to a point process with a specified intensity function. In order to do this, we need to calculate the maximum possible intensity over the region, based on our pre-specified intensity function that we are simulating from. For the two-stage model, the marks are incorporated in the last step of the simulation process. For the bivariate mark model, the intensity involves mark-dependent parameters, so the maximum intensity must be calculated for each mark. Another key difference between the bivariate mark and two-stage simulation procedures is that in the bivariate mark model, the marks and nonspatial variables are simulated before thinning, where in the two-stage models, these variables are assigned after thinning.

The bivariate marked point process model has parameters $\alpha_1, \dots, \alpha_p$ for each mark $k$ ($\bm{\alpha_k})$, the regression coefficients for the nonspatial variables, $\beta_1,\dots, \beta_k$ for each mark $k$ ($\bm{\beta_k})$, the coefficients for the spatial variables, and covariance parameters $\sigma_1, \sigma_2, >0 $ and $\rho \in (-1,1)$. As shown in Section \ref{sec:bivmodel_sim}, we need to simulate nonspatial variables from uniform distributions/equal probability distributions for the bivariate mark model.

\begin{algorithm}
  \caption{Bivariate Marked Point Process Model Simulation}
  \label{bivar_algorithm}
  \begin{algorithmic}[1]
    \State Define $n$ locations, $x_1, \dots, x_n \in W^\dag$. Simulate a bivariate Gaussian process, $\omega_1(x_1),\dots,\omega_1(x_n)$ and $\omega_2(x_1),\dots,\omega_2(x_n)$ with parameters $\sigma_1, \sigma_2,$ and $\rho$.
    \State Decide the bounds of continuous nonspatial variables from their distributions in order to calculate the maximum contribution to the intensity function by the continuous nonspatial variables, $\max(\bm{\nu}'\bm{\alpha})$.
    \State \textbf{Calculate maximum intensity} for each mark, $\max(\lambda_{\bm{\theta}_1}(s))$ and $\max(\lambda_{\bm{\theta}_2}(s))$.

    \State \textbf{Simulate a homogeneous Poisson Process for each mark}, with parameter $\max(\lambda_{\bm{\theta}_k}(s))\times a$ for mark $k$, on $W$, where $a$ is the area of $W$.
    \State \textbf{Simulate nonspatial variables} according to their distributions and assign values of the nonspatial variables to each point from the unthinned homogeneous Poisson process. 
    \State Affiliate each simulated point with one of $x_1, \dots, x_n$ \hspace{-5pt}$^\ddag$. This is to assign a value of $\omega_1(s)$ and $\omega_2(s)$ to each simulated point.
    \State \textbf{Thin the points} by keeping the location $s$ of mark $k$ in the homogeneous Poisson process with probability $\lambda_{\bm{\theta}_k}(s)/\max(\lambda_{\bm{\theta}_k}(s))$, otherwise remove.
    
    \vspace{12pt}
    \hspace{-40pt} $\dag$: In our approach, we found it helpful to generate $x_1,\dots x_n$ based on a grid of equally spaced points in the observation window, $W$.
    
    \hspace{-40pt} $\ddag$: Options for affiliating each simulated point with $x_1,\dots,x_n$ include manual affiliation with the closest point or predictive process transformations.
  \end{algorithmic}
\end{algorithm}

\section{Posterior Mean Estimates, Models 1-4, Forest Fire Data}

Table \ref{tab:tab05_full} includes results for models 1-4, as applied to the forest fire data from the Castilla-La Mancha region of Spain.

\begin{table}
 %\\
\centering
\begin{tabular}{|llrr|}
\hline
& & NHPP-IndNorm (Mod 1) & UniGP-IndNorm (Mod 2) \\
\hline
&Variable & Mean (95\%CI) & Mean (95\%CI)  \\ 
 \hline
 \multirow{5}{*}{\rotatebox[origin=c]{90}{Stage 1}} &$\beta_0$ (intercept) & {\bf4.4404} (4.3208, 4.5586) & {\bf4.3464} (4.0826, 4.6114) \\ 
 &Forest (ind.) & 0.0793 (-0.0189, 0.1801) &  {\bf-0.1992} (-0.3091, -0.091)  \\ 
 &Elevation & {\bf-9${\bm{\times10^{-04}}}$} (-0.0011, -8$\times10^{-04}$) & {\bf-0.0012} (-0.0015, -9$\times10^{-04}$) \\ 
 &Slope & {\bf0.0283} (0.0219, 0.0346) & {\bf0.0242} (0.0168, 0.0317)\\ 
&$\sigma_1$ &   & {\bf5.2065} (4.6201, 5.7915)  \\ 
 \hline
 \multirow{7}{*}{\rotatebox[origin=c]{90}{ Stage 2}}& $\alpha_0$ (intercept) &{\bf1.4706} (1.1702, 1.7756) & {\bf1.477} (1.1699, 1.7816) \\ 
 &Intentional (ind.) & {\bf0.2357} (0.0644, 0.4019) & {\bf0.2343} (0.0646, 0.4052) \\ 
 &Summer (ind.) &{\bf-0.1563} (-0.3105, -0.0026) & {\bf-0.1589} (-0.3125, -0.0083) \\ 
 &Forest (ind.) &{\bf-0.7141 (-0.9498, -0.4789)} & {\bf-0.7122} (-0.9447, -0.4732) \\ 
 &Elevation & {\bf-0.0027} (-0.0031, -0.0024)&  {\bf-0.0027} (-0.0031, -0.0023)  \\ 
 &Slope & {\bf-0.0162} (-0.0322, -1$\times10^{-04}$) & {\bf-0.0164} (-0.0326, -1$\times10^{-04}$) \\ 
  & $\sigma$ (iid) & {\bf2.3238} (2.2709, 2.378) & {\bf2.3235} (2.2714, 2.3774)  \\
\hline

\hline
& & UniGP-UniGP (Mod 3) & BivGP (Mod 4)\\
\hline
&Variable & Mean (95\%CI) & Mean (95\%CI)  \\ 
 \hline
 \multirow{5}{*}{\rotatebox[origin=c]{90}{Stage 1}} &$\beta_0$ (intercept) & {\bf4.492} (4.3008, 4.6831) & {\bf4.0914} (3.888, 4.2874) \\ 
 &Forest (ind.) & {\bf-0.1929} (-0.3005, -0.0825)& {\bf-0.1947} (-0.299, -0.0835)\\ 
 &Elevation &{\bf-0.0013} (-0.0015, -0.0011) & {\bf-0.001} (-0.0012, -7$\times10^{-04}$)\\ 
 &Slope & {\bf0.026} (0.0186, 0.033)& {\bf0.0252} (0.0179, 0.0322) \\ 
&$\sigma_1$ &  {\bf5.1529} (4.6399, 5.6641) & {\bf5.3198} (4.8447, 5.7995) \\ 
 \hline
 \multirow{7}{*}{\rotatebox[origin=c]{90}{ Stage 2}}& $\alpha_0$ (intercept) & {\bf1.4333} (1.0825, 1.7868)& {\bf1.4759} (0.93, 1.9269)\\ 
 &Intentional (ind.) & 0.0286 (-0.1358, 0.191)& 0.0607 (-0.1049, 0.2287)\\ 
 &Summer (ind.) & {\bf-0.2191} (-0.3678, -0.0747) & {\bf-0.2124} (-0.3635, -0.0664) \\ 
 &Forest (ind.) & {\bf-0.4988} (-0.7337, -0.2745) & {\bf-0.4572} (-0.694, -0.2267) \\ 
 &Elevation & {\bf-0.0027} (-0.0031, -0.0022) & {\bf-0.0025} (-0.003, -0.0019) \\ 
 &Slope & 0.0074 (-0.0088, 0.0228)& -0.0036 (-0.0208, 0.0133) \\ 
  & $\sigma$ (iid) & {\bf2.1906} (2.1399, 2.2416) & {\bf2.2135} (2.1596, 2.2698)\\
  &$\sigma_2$ & {\bf3.4521} (2.7675, 4.2067) & {\bf6.3139} (5.6398, 7.0931) \\
 \hline
& $\rho$ & & {\bf-0.1435} (-0.2598, -0.024) \\
\hline
\end{tabular}
\caption{\label{tab:tab05_full} We present the posterior mean as well as the 95\% Credible Intervals for all covariates in our analysis of the Spain Fire data in the Castilla-La Mancha region of Spain. Values in bold indicate coefficients whose credible intervals do not contain 0.}
\end{table}

\bibliography{references}

\begin{thebibliography}{28}
\expandafter\ifx\csname natexlab\endcsname\relax\def\natexlab#1{#1}\fi
\providecommand{\url}[1]{\texttt{#1}}
\providecommand{\href}[2]{#2}
\providecommand{\path}[1]{#1}
\providecommand{\DOIprefix}{doi:}
\providecommand{\ArXivprefix}{arXiv:}
\providecommand{\URLprefix}{URL: }
\providecommand{\Pubmedprefix}{pmid:}
\providecommand{\doi}[1]{\href{http://dx.doi.org/#1}{\path{#1}}}
\providecommand{\Pubmed}[1]{\href{pmid:#1}{\path{#1}}}
\providecommand{\bibinfo}[2]{#2}
\ifx\xfnm\relax \def\xfnm[#1]{\unskip,\space#1}\fi
%Type = Article
\bibitem[{Alba-Fern{\'a}ndez and Ariza-L{\'o}pez(2018)}]{alba2018homogeneity}
\bibinfo{author}{Alba-Fern{\'a}ndez, M.V.}, \bibinfo{author}{Ariza-L{\'o}pez,
  F.J.}, \bibinfo{year}{2018}.
\newblock \bibinfo{title}{A homogeneity test for comparing
  gridded-spatial-point patterns of human caused fires}.
\newblock \bibinfo{journal}{Forests} \bibinfo{volume}{9}, \bibinfo{pages}{454}.
%Type = Book
\bibitem[{Baddeley et~al.(2015)Baddeley, Rubak and Turner}]{spatstat}
\bibinfo{author}{Baddeley, A.}, \bibinfo{author}{Rubak, E.},
  \bibinfo{author}{Turner, R.}, \bibinfo{year}{2015}.
\newblock \bibinfo{title}{Spatial Point Patterns: Methodology and Applications
  with {R}}.
\newblock \bibinfo{publisher}{Chapman and Hall/CRC Press},
  \bibinfo{address}{London}.
\newblock \URLprefix
  \url{http://www.crcpress.com/Spatial-Point-Patterns-Methodology-and-Applications-with-R/Baddeley-Rubak-Turner/9781482210200/}.
%Type = Book
\bibitem[{Banerjee et~al.(2014)Banerjee, Carlin and
  Gelfand}]{banerjee2014hierarchical}
\bibinfo{author}{Banerjee, S.}, \bibinfo{author}{Carlin, B.P.},
  \bibinfo{author}{Gelfand, A.E.}, \bibinfo{year}{2014}.
\newblock \bibinfo{title}{Hierarchical modeling and analysis for spatial data}.
\newblock \bibinfo{publisher}{Crc Press}.
%Type = Article
\bibitem[{Banerjee et~al.(2008)Banerjee, Gelfand, Finley and
  Sang}]{banerjee2008gaussian}
\bibinfo{author}{Banerjee, S.}, \bibinfo{author}{Gelfand, A.E.},
  \bibinfo{author}{Finley, A.O.}, \bibinfo{author}{Sang, H.},
  \bibinfo{year}{2008}.
\newblock \bibinfo{title}{Gaussian predictive process models for large spatial
  data sets}.
\newblock \bibinfo{journal}{Journal of the Royal Statistical Society: Series B
  (Statistical Methodology)} \bibinfo{volume}{70}, \bibinfo{pages}{825--848}.
%Type = Article
\bibitem[{Best et~al.(2000)Best, Ickstadt and Wolpert}]{best2000spatial}
\bibinfo{author}{Best, N.G.}, \bibinfo{author}{Ickstadt, K.},
  \bibinfo{author}{Wolpert, R.L.}, \bibinfo{year}{2000}.
\newblock \bibinfo{title}{Spatial {P}oisson regression for health and exposure
  data measured at disparate resolutions}.
\newblock \bibinfo{journal}{Journal of the American {S}tatistical
  {A}ssociation} \bibinfo{volume}{95}, \bibinfo{pages}{1076--1088}.
%Type = Article
\bibitem[{Brix and M{\o}ller(2001)}]{brix2001space}
\bibinfo{author}{Brix, A.}, \bibinfo{author}{M{\o}ller, J.},
  \bibinfo{year}{2001}.
\newblock \bibinfo{title}{Space-time multi type log {G}aussian {C}ox processes
  with a view to modelling weeds}.
\newblock \bibinfo{journal}{Scandinavian Journal of Statistics}
  \bibinfo{volume}{28}, \bibinfo{pages}{471--488}.
%Type = Article
\bibitem[{{de Valpine} et~al.(2017){de Valpine}, Turek, Paciorek,
  Anderson-Bergman, {Temple Lang} and Bodik}]{nimble}
\bibinfo{author}{{de Valpine}, P.}, \bibinfo{author}{Turek, D.},
  \bibinfo{author}{Paciorek, C.}, \bibinfo{author}{Anderson-Bergman, C.},
  \bibinfo{author}{{Temple Lang}, D.}, \bibinfo{author}{Bodik, R.},
  \bibinfo{year}{2017}.
\newblock \bibinfo{title}{Programming with models: writing statistical
  algorithms for general model structures with {NIMBLE}}.
\newblock \bibinfo{journal}{Journal of Computational and Graphical Statistics}
  \bibinfo{volume}{26}, \bibinfo{pages}{403--413}.
\newblock \DOIprefix\doi{10.1080/10618600.2016.1172487}.
%Type = Article
\bibitem[{Diggle et~al.(2010)Diggle, Guan, Hart, Paize and
  Stanton}]{diggle2010estimating}
\bibinfo{author}{Diggle, P.J.}, \bibinfo{author}{Guan, Y.},
  \bibinfo{author}{Hart, A.C.}, \bibinfo{author}{Paize, F.},
  \bibinfo{author}{Stanton, M.}, \bibinfo{year}{2010}.
\newblock \bibinfo{title}{Estimating individual-level risk in spatial
  epidemiology using spatially aggregated information on the population at
  risk}.
\newblock \bibinfo{journal}{Journal of the American Statistical Association}
  \bibinfo{volume}{105}, \bibinfo{pages}{1394--1402}.
%Type = Article
\bibitem[{Dvo{\v{r}}{\'a}k et~al.(2020)Dvo{\v{r}}{\'a}k, Mrkvi{\v{c}}ka, Mateu
  and Gonz{\'a}lez}]{dvovrak2020nonparametric}
\bibinfo{author}{Dvo{\v{r}}{\'a}k, J.}, \bibinfo{author}{Mrkvi{\v{c}}ka, T.},
  \bibinfo{author}{Mateu, J.}, \bibinfo{author}{Gonz{\'a}lez, J.},
  \bibinfo{year}{2020}.
\newblock \bibinfo{title}{Nonparametric testing of the dependence structure
  among points-marks-covariates in spatial point patterns}.
\newblock \bibinfo{journal}{arXiv preprint arXiv:2005.01019} .
%Type = Manual
\bibitem[{Flegal et~al.(2021)Flegal, Hughes, Vats, Dai, Gupta and
  Maji}]{mcmcse_package}
\bibinfo{author}{Flegal, J.M.}, \bibinfo{author}{Hughes, J.},
  \bibinfo{author}{Vats, D.}, \bibinfo{author}{Dai, N.},
  \bibinfo{author}{Gupta, K.}, \bibinfo{author}{Maji, U.},
  \bibinfo{year}{2021}.
\newblock \bibinfo{title}{mcmcse: {M}onte {C}arlo Standard Errors for {MCMC}}.
\newblock \bibinfo{address}{Riverside, CA, and Kanpur, India}.
\newblock \bibinfo{note}{R package version 1.5-0}.
%Type = Article
\bibitem[{Guan(2006)}]{guan2006tests}
\bibinfo{author}{Guan, Y.}, \bibinfo{year}{2006}.
\newblock \bibinfo{title}{Tests for independence between marks and points of a
  marked point process}.
\newblock \bibinfo{journal}{Biometrics} \bibinfo{volume}{62},
  \bibinfo{pages}{126--134}.
%Type = Manual
\bibitem[{Haran and Hughes(2020)}]{batchmeans_package}
\bibinfo{author}{Haran, M.}, \bibinfo{author}{Hughes, J.},
  \bibinfo{year}{2020}.
\newblock \bibinfo{title}{batchmeans: {C}onsistent Batch Means Estimation of
  {M}onte {C}arlo Standard Errors}.
\newblock \bibinfo{address}{Frederick, MD}.
\newblock \bibinfo{note}{R package version 1.0-4}.
%Type = Article
\bibitem[{Li et~al.(2012)Li, Brown, Gesink and Rue}]{li2012log}
\bibinfo{author}{Li, Y.}, \bibinfo{author}{Brown, P.}, \bibinfo{author}{Gesink,
  D.C.}, \bibinfo{author}{Rue, H.}, \bibinfo{year}{2012}.
\newblock \bibinfo{title}{Log {G}aussian {C}ox processes and spatially
  aggregated disease incidence data}.
\newblock \bibinfo{journal}{Statistical methods in medical research}
  \bibinfo{volume}{21}, \bibinfo{pages}{479--507}.
%Type = Article
\bibitem[{Liang et~al.(2008)Liang, Carlin and Gelfand}]{liang2008analysis}
\bibinfo{author}{Liang, S.}, \bibinfo{author}{Carlin, B.P.},
  \bibinfo{author}{Gelfand, A.E.}, \bibinfo{year}{2008}.
\newblock \bibinfo{title}{Analysis of {M}innesota colon and rectum cancer point
  patterns with spatial and nonspatial covariate information}.
\newblock \bibinfo{journal}{The Annals of Applied Statistics}
  \bibinfo{volume}{3}, \bibinfo{pages}{943}.
%Type = Article
\bibitem[{Michaud et~al.(2014)Michaud, Coops, Andrew, Wulder, Brown and
  Rickbeil}]{michaud2014estimating}
\bibinfo{author}{Michaud, J.S.}, \bibinfo{author}{Coops, N.C.},
  \bibinfo{author}{Andrew, M.E.}, \bibinfo{author}{Wulder, M.A.},
  \bibinfo{author}{Brown, G.S.}, \bibinfo{author}{Rickbeil, G.J.},
  \bibinfo{year}{2014}.
\newblock \bibinfo{title}{Estimating moose ({A}lces alces) occurrence and
  abundance from remotely derived environmental indicators}.
\newblock \bibinfo{journal}{Remote Sensing of Environment}
  \bibinfo{volume}{152}, \bibinfo{pages}{190--201}.
%Type = Article
\bibitem[{M{\o}ller et~al.(1998)M{\o}ller, Syversveen and
  Waagepetersen}]{moller1998log}
\bibinfo{author}{M{\o}ller, J.}, \bibinfo{author}{Syversveen, A.R.},
  \bibinfo{author}{Waagepetersen, R.P.}, \bibinfo{year}{1998}.
\newblock \bibinfo{title}{Log {G}aussian {C}ox processes}.
\newblock \bibinfo{journal}{Scandinavian Journal of Statistics}
  \bibinfo{volume}{25}, \bibinfo{pages}{451--482}.
%Type = Book
\bibitem[{M{\o}ller and Waagepetersen(2003)}]{moller2003statistical}
\bibinfo{author}{M{\o}ller, J.}, \bibinfo{author}{Waagepetersen, R.P.},
  \bibinfo{year}{2003}.
\newblock \bibinfo{title}{Statistical inference and simulation for spatial
  point processes}.
\newblock \bibinfo{publisher}{CRC Press}.
%Type = Article
\bibitem[{Myllym{\"a}ki et~al.(2020)Myllym{\"a}ki, Kuronen and
  Mrkvi{\v{c}}ka}]{myllymaki2020testing}
\bibinfo{author}{Myllym{\"a}ki, M.}, \bibinfo{author}{Kuronen, M.},
  \bibinfo{author}{Mrkvi{\v{c}}ka, T.}, \bibinfo{year}{2020}.
\newblock \bibinfo{title}{Testing global and local dependence of point patterns
  on covariates in parametric models}.
\newblock \bibinfo{journal}{Spatial Statistics} , \bibinfo{pages}{100436}.
%Type = Article
\bibitem[{Myllym{\"a}ki and Penttinen(2009)}]{myllymaki2009conditionally}
\bibinfo{author}{Myllym{\"a}ki, M.}, \bibinfo{author}{Penttinen, A.},
  \bibinfo{year}{2009}.
\newblock \bibinfo{title}{Conditionally heteroscedastic intensity-dependent
  marking of log gaussian cox processes}.
\newblock \bibinfo{journal}{Statistica Neerlandica} \bibinfo{volume}{63},
  \bibinfo{pages}{450--473}.
%Type = Article
\bibitem[{Pinto~Junior et~al.(2015)Pinto~Junior, Gamerman, Paez and
  Fonseca~Alves}]{pinto2015point}
\bibinfo{author}{Pinto~Junior, J.A.}, \bibinfo{author}{Gamerman, D.},
  \bibinfo{author}{Paez, M.S.}, \bibinfo{author}{Fonseca~Alves, R.H.},
  \bibinfo{year}{2015}.
\newblock \bibinfo{title}{Point pattern analysis with spatially varying
  covariate effects, applied to the study of cerebrovascular deaths}.
\newblock \bibinfo{journal}{Statistics in Medicine} \bibinfo{volume}{34},
  \bibinfo{pages}{1214--1226}.
%Type = Article
\bibitem[{Quick et~al.(2015)Quick, Holan, Wikle and Reiter}]{quick2015bayesian}
\bibinfo{author}{Quick, H.}, \bibinfo{author}{Holan, S.H.},
  \bibinfo{author}{Wikle, C.K.}, \bibinfo{author}{Reiter, J.P.},
  \bibinfo{year}{2015}.
\newblock \bibinfo{title}{Bayesian marked point process modeling for generating
  fully synthetic public use data with point-referenced geography}.
\newblock \bibinfo{journal}{Spatial Statistics} \bibinfo{volume}{14},
  \bibinfo{pages}{439--451}.
%Type = Article
\bibitem[{Recta et~al.(2012)Recta, Haran and Rosenberger}]{recta2012two}
\bibinfo{author}{Recta, V.}, \bibinfo{author}{Haran, M.},
  \bibinfo{author}{Rosenberger, J.L.}, \bibinfo{year}{2012}.
\newblock \bibinfo{title}{A two-stage model for incidence and prevalence in
  point-level spatial count data}.
\newblock \bibinfo{journal}{Environmetrics} \bibinfo{volume}{23},
  \bibinfo{pages}{162--174}.
%Type = Article
\bibitem[{Schlather et~al.(2004)Schlather, Ribeiro~Jr and
  Diggle}]{schlather2004detecting}
\bibinfo{author}{Schlather, M.}, \bibinfo{author}{Ribeiro~Jr, P.J.},
  \bibinfo{author}{Diggle, P.J.}, \bibinfo{year}{2004}.
\newblock \bibinfo{title}{Detecting dependence between marks and locations of
  marked point processes}.
\newblock \bibinfo{journal}{Journal of the Royal Statistical Society: Series B
  (Statistical Methodology)} \bibinfo{volume}{66}, \bibinfo{pages}{79--93}.
%Type = Article
\bibitem[{Shirota and Gelfand(2017)}]{shirota2017space}
\bibinfo{author}{Shirota, S.}, \bibinfo{author}{Gelfand, A.E.},
  \bibinfo{year}{2017}.
\newblock \bibinfo{title}{Space and circular time log {G}aussian {C}ox
  processes with application to crime event data}.
\newblock \bibinfo{journal}{The Annals of Applied Statistics} ,
  \bibinfo{pages}{481--503}.
%Type = Article
\bibitem[{Siino et~al.(2018)Siino, Adelfio and Mateu}]{siino2018joint}
\bibinfo{author}{Siino, M.}, \bibinfo{author}{Adelfio, G.},
  \bibinfo{author}{Mateu, J.}, \bibinfo{year}{2018}.
\newblock \bibinfo{title}{Joint second-order parameter estimation for
  spatio-temporal log-{G}aussian {C}ox processes}.
\newblock \bibinfo{journal}{Stochastic environmental research and risk
  assessment} \bibinfo{volume}{32}, \bibinfo{pages}{3525--3539}.
%Type = Article
\bibitem[{Ver~Hoef and Jansen(2007)}]{ver2007space}
\bibinfo{author}{Ver~Hoef, J.M.}, \bibinfo{author}{Jansen, J.K.},
  \bibinfo{year}{2007}.
\newblock \bibinfo{title}{Space-time zero-inflated count models of {H}arbor
  seals}.
\newblock \bibinfo{journal}{Environmetrics: The official journal of the
  International Environmetrics Society} \bibinfo{volume}{18},
  \bibinfo{pages}{697--712}.
%Type = Article
\bibitem[{Waagepetersen et~al.(2016)Waagepetersen, Guan, Jalilian and
  Mateu}]{waagepetersen2016analysis}
\bibinfo{author}{Waagepetersen, R.}, \bibinfo{author}{Guan, Y.},
  \bibinfo{author}{Jalilian, A.}, \bibinfo{author}{Mateu, J.},
  \bibinfo{year}{2016}.
\newblock \bibinfo{title}{Analysis of multispecies point patterns by using
  multivariate log-{G}aussian {C}ox processes}.
\newblock \bibinfo{journal}{Journal of the Royal Statistical Society: Series C:
  Applied Statistics} , \bibinfo{pages}{77--96}.
%Type = Article
\bibitem[{Watanabe and Opper(2010)}]{watanabe2010asymptotic}
\bibinfo{author}{Watanabe, S.}, \bibinfo{author}{Opper, M.},
  \bibinfo{year}{2010}.
\newblock \bibinfo{title}{Asymptotic equivalence of {B}ayes cross validation
  and widely applicable information criterion in singular learning theory.}
\newblock \bibinfo{journal}{Journal of Machine Learning Research}
  \bibinfo{volume}{11}.

\end{thebibliography}

\end{document}